\documentclass{aastex631}

\usepackage{amsthm,amsmath,amssymb}
\usepackage{lipsum}
\usepackage{float}
\usepackage{xcolor}
\usepackage{soul}

\begin{document}

\title{BSN-IV: The First Multiband Light Curve Study of Five W UMa-type Contact Binary Systems}

\author[0000-0002-0196-9732]{Atila Poro}
\altaffiliation{atila.poro@obspm.fr, atilaporo@bsnp.info (AP)}
\affiliation{LUX, Observatoire de Paris, CNRS, PSL, 61 Avenue de l'Observatoire, 75014 Paris, France}
\affiliation{Astronomy Department of the Raderon AI Lab., BC., Burnaby, Canada}

\author[0000-0003-1263-808X]{Raul Michel}
\altaffiliation{rmm@astro.unam.mx (RM)}
\affiliation{Instituto de Astronom\'ia, UNAM. A.P. 106, 22800 Ensenada, BC, M\'exico}

\author[0000-0003-0354-8568]{Jean-François Coliac}
\affiliation{Double Stars Committee, Société Astronomique de France, Paris, France}

\author[0009-0006-9150-3392]{Maryam Nastaran}
\affiliation{Faculty of Geology, University of Tehran, 1417935840 Tehran, Iran}

\author[0000-0002-9262-4456]{Eduardo Fernández Lajús}
\affiliation{Instituto de Astrofísica de La Plata (CCT La Plata-CONICET-UNLP), La Plata, Argentina}

\author[0000-0002-9761-9509]{Francisco Javier Tamayo}
\affiliation{Facultad de Ciencias F\'{\i}sico-Matem\'aticas, UANL, 66451 San Nicol\'as de los Garza, NL, M\'exico}

\author[0000-0002-7348-8815]{Hector Aceves}
\affiliation{Instituto de Astronom\'ia, UNAM. A.P. 106, 22800 Ensenada, BC, M\'exico}

\author[0000-0002-1972-8400]{Fahri Alicavus}
\affiliation{Çanakkale Onsekiz Mart University, Faculty of Arts and Sciences, Department of Physics, 17020, Çanakkale, Türkiye}
\affiliation{Çanakkale Onsekiz Mart University, Astrophysics Research Center and Ulupnar Observatory, 17020, Çanakkale,
Türkiye}

\author{Morgan-Rhai Najera}
\affiliation{Instituto de Astronom\'ia, UNAM. A.P. 106, 22800 Ensenada, BC, M\'exico}

\begin{abstract}
In this work, we present a detailed investigation of five contact binary systems of the W Ursae Majoris (W UMa) type. Multiband photometric observations were conducted using ground-based telescopes in both the northern and southern hemispheres, yielding new times of minima. O–C diagram analysis reveals that two systems exhibit parabolic trends, indicating a gradual long-term decrease in their orbital periods. The light curves were modeled using version 1.0 of the BSN application, with one system requiring the inclusion of a cool starspot to achieve a satisfactory fit. We examined empirical relationships between orbital period and fundamental parameters, identifying the period–semi-major axis ($P$–$a$) relation as the most robust correlation, which was used to estimate absolute parameters. To statistically assess thermal equilibrium, we analyzed temperature differences between components and found that 90\% of systems exhibit less than 9.4\% contrast. Two target systems with extremely low mass ratios were identified, and their orbital stability was evaluated. Based on the effective temperatures and component masses, two systems were classified as W-subtype and three as A-subtype. The evolutionary status of the binaries was assessed through their locations in mass–radius, mass–luminosity, and other empirical diagrams, and initial component masses as well as total mass loss were also estimated.
\end{abstract}


\section{Introduction}
Contact binaries are an important class of close binary stars, consisting of two stellar components that share a common envelope, enabling the exchange of both mass and energy between them (\citealt{1968ApJ...151.1123L}, \citealt{1968ApJ...153..877L}). These systems are predominantly composed of late-type stars, including spectral types F, G, K, and occasionally M, and they represent a considerable portion of the stellar population. Estimates suggest that roughly one in every 500 main-sequence stars in our Galaxy is part of a contact binary (\citealt{2007MNRAS.382..393R}). Contact binaries are further categorized into A-subtype and W-subtype systems, distinguished by which component is hotter: the more massive star in A-subtype systems or the less massive one in W-subtype systems (\citealt{1970VA.....12..217B}). Despite substantial research, the evolutionary connections between these subtypes remain a matter of debate (\citealt{2020MNRAS.492.4112Z}).

A widely accepted scenario proposes that contact binaries originate from short-period detached binaries that evolve into contact configurations through angular momentum loss driven by magnetic braking (\citealt{1982AA...109...17V}, \citealt{2006AcA....56..347S}, \citealt{2017RAA....17...87Q}). Over time, these systems may merge into rapidly rotating single stars, potentially giving rise to objects like blue stragglers or FK Com-type stars (\citealt{1995ApJ...444L..41R}). However, direct observational evidence for such mergers is limited, with V1309 Sco being one of the few well-documented examples (\citealt{2011AA...528A.114T}).

The distribution of orbital periods among contact binaries typically peaks near 0.27 days, while a well-established short-period cutoff around 0.22 days suggests a physical limit potentially imposed by angular momentum loss processes or structural instabilities in low-mass components (\citealt{2012MNRAS.421.2769J}). Moreover, variations in orbital periods are strongly linked to mass transfer dynamics and angular momentum loss mechanisms such as magnetic braking (\citealt{2013ApJS..209...13Q}).

A characteristic phenomenon observed in many contact binaries is the O'Connell effect, where the two maxima in their light curves show unequal brightness. This asymmetry is often linked to magnetic activity and starspots on the stellar surfaces, though other explanations such as hot spots caused by mass transfer or the presence of circumstellar material have also been suggested (\citealt{1951PRCO....2...85O}, \citealt{2003ChJAA...3..142L}). Clarifying the origin of this effect is crucial for accurately determining the physical parameters of these systems, as surface inhomogeneities can significantly impact photometric analyses (\citealt{2021PASP..133h4202L}).

In this work, ground-based multiband photometric observations were conducted for five eclipsing contact binary systems of the W UMa type, aiming to refine their orbital and physical parameter estimates. This research extends the efforts begun by \cite{2025MNRAS.537.3160P, 2025AJ....170..214P, 2025PASP..137h4201P}, offering new observations and analyses focused on additional W UMa-type contact binary stars within the framework of the BSN project\footnote{\url{https://bsnp.info/}}. The paper is structured as follows: Section 2 introduces the target systems. Section 3 describes the acquisition and reduction of both ground- and space-based photometric data. Variations in orbital periods are analyzed in Section 4. Section 5 presents the solutions to the photometric light curves of the targets. The determination of absolute parameters is detailed in Section 6. Finally, Section 7 discusses the findings and summarizes the conclusions.

\vspace{0.6cm}
\section{Target Binary Stars}
We have analyzed five eclipsing binary stars, including BF Dor, CRTS J014306.6+383909 (hereinafter J014306), LZ Leo, WISE J201317.7+464036 (hereinafter J201317), and ZTF J004331.79+500059.9 (hereinafter J004331). These contact binary systems had not been studied in detail previously. Also, we had multiband photometric data available in the BSN project database for these systems, providing sufficient observational coverage for accurate analysis. Table \ref{systemsinfo} presents specifications for the target systems based on the Gaia DR3 database (\citealt{2023AA...674A..33G}). The general properties of the target systems are summarized below:

$\bullet$ BF Dor: According to the Variable Star Index (VSX) report, this system was discovered by Friedhelm Hund. In both the VSX database and the General Catalog of Variable Stars (GCVS; \citealt{2017ARep...61...80S}), BF Dor is classified as a contact binary system with an orbital period of 0.352271 days. However, in the All-Sky Automated Survey for Supernovae (ASAS-SN; \citealt{2018MNRAS.477.3145J}) and the Zwicky Transient Facility (ZTF; \citealt{2023AA...675A.195S}) catalogs, it is listed as an EA-type (detached) system with a period of 0.70455 days; twice the value of GCVS. The VSX reports a maximum apparent magnitude of $13^{\text{mag.}}$ for this system.

$\bullet$ J014306: This binary system was discovered in the Catalina Surveys Data Release 1 (CSDR1; \citealt{2014ApJS..213....9D}). The CSDR1, VSX, ZTF, and ASAS-SN catalogs consistently classify J014306 as a contact binary system with an orbital period of 0.27923 days. The VSX database reports an apparent magnitude of $14.30^{\text{mag.}}$ for this system.

$\bullet$ LZ Leo: This system was discovered in the CSDR1 catalog. LZ Leo is classified as a contact binary star in the VSX, ASAS-SN, ZTF, and GCVS catalogs. The VSX and GCVS databases report maximum apparent magnitudes of $13.19^{\text{mag.}}$ and $12.97^{\text{mag.}}$ for this system, respectively.

$\bullet$ J201317: This system is classified as a contact binary in the Wide-field Infrared Survey Explorer (WISE; \citealt{2018ApJS..237...28C}) catalog. Limited information about J201317 is available in existing catalogs. Its orbital period is listed as 0.2967596 days in the VSX database. The effective temperature reported in the TESS Input Catalog (TIC) is $5143\pm195$ K, which is consistent with the temperature derived from Gaia DR3.

$\bullet$ J004331: This system was discovered by the Asteroid Terrestrial-impact Last Alert System (ATLAS; \citealt{2018ApJ...867..105T}) catalog. J004331 is reported as a contact binary in the VSX, ASAS-SN, and ZTF catalogs, with an orbital period of 0.392775 days. The VSX database lists its apparent magnitude as $13.373(168)^{\text{mag.}}$ in $r$ passband.

\begin{table*}
\renewcommand\arraystretch{1.2}
\caption{Specifications of the target systems from Gaia DR3.}
\centering
\begin{center}
\footnotesize
\begin{tabular}{c c c c c c}
\hline
System & RA$.^\circ$(J2000) & Dec$.^\circ$(J2000) & $d$(pc) & RUWE & $T$(K)\\
\hline
BF Dor	&	91.477281	&	-66.843424	&	615(6)	&	1.146	&	5636(10)\\
CRTS J014306.6+383909 (J014306)	&	25.778003	&	38.652536	&	525(6)	&	1.022	&	4921(6)\\
LZ Leo	&	149.339006	&	14.204328	&	403(3)	&	1.000	&	-\\
WISE J201317.7+464036 (J201317)	&	303.323961	&	46.676887	&	943(21)	&	1.014	&	5259(46)\\
ZTF J004331.79+500059.9 (J004331)	&	10.882479	&	50.016630	&	1349(47)	&	2.135	&	5259(19)\\
\hline
\end{tabular}
\end{center}
\label{systemsinfo}
\end{table*}

\vspace{0.6cm}
\section{Observation and Data Reduction}
Photometric observations and data reduction for the five target binary systems were carried out using standard filters at three observatories located in both the Northern and Southern Hemispheres: Complejo Astronómico El Leoncito (CASLEO), San Pedro Mártir (SPM), and the Observatoire Astronomique des Binaires André Coliac (OABAC). Table \ref{observations} provides details for each system, including the observation date, type of filter, exposure time (in seconds), the maximum apparent magnitude of the systems in our observations in filter $V$, and the observatory where the data were obtained. The general characteristics of the comparison and check stars used during the observations and data reductions are shown in Table \ref{stars}. These stars played a critical role in ensuring the precision and stability of the photometric measurements. The use of comparison and check stars helped minimize systematic errors and improve the accuracy of the resulting light curves.

\subsection{CASLEO Observatory}
Observations of BF Dor were carried out with the 2.15-meter Jorge Sahade (JS) telescope at the CASLEO Observatory in Argentina (located at $69^\circ 18'$ W, $31^\circ 48'$ S, 2552 m elevation). Data acquisition employed a Versarray 2048B CCD camera (Roper Scientific, Princeton Instruments) in combination with standard $BVR_cI_c$ filters. The detector provided a plate scale of 0.15 arcsec/pixel, and observations were performed using a $5 \times 5$ binning mode. CCD data reduction and aperture photometry were conducted with the APPHOT package within the Image Reduction and Analysis Facility (IRAF, \citealt{1986SPIE..627..733T}), utilizing bias frames and flat-field corrections.

\vspace{0.3cm}
\subsection{SPM Observatory}
The binary systems J014306 and J201317 were observed at the SPM Observatory in México, situated at longitude $115^\circ$ $27'$ $49''$ W, latitude $31^\circ$ $02'$ $39''$ N, and an altitude of 2830 meters. Observations were performed using two Ritchey-Chrétien telescopes. The 0.84-meter telescope, operating at an $f/15$ focal ratio, was equipped with the Mexman filter wheel and a Marconi 5 CCD camera (e2v CCD231-42), featuring $15 \times 15 , \mu\mathrm{m}^2$ pixels, a gain of $2.2 , e^- /\mathrm{ADU}$, and a readout noise of $3.6 , e^-$. The 1.5-meter telescope employed the RUCA filter wheel along with the Spectral Instruments 1 detector, housing an e2v CCD42-40 chip with $13.5 \times 13.5 , \mu\mathrm{m}^2$ pixels, a gain of $1.39 , e^- /\mathrm{ADU}$, and a readout noise of $3.49 , e^-$. Standard $B$, $V$, $R_c$, and $I_c$ filters were used for these observations. Data reduction and photometry were carried out with IRAF software tools, following standard procedures such as bias subtraction and flat-field correction, as described by \cite{1986SPIE..627..733T}.

\vspace{0.3cm}
\subsection{OABAC Observatory}
The OABAC in Marseille, France, located at longitude $05^\circ$ $27'$ $56''$ E and latitude $43^\circ$ $18'$ $54''$ N. For LZ Leo, a 200 mm Newtonian telescope with an $f/4$ focal ratio was used, paired with an ASI ZWO 533MM Pro CCD camera and standard $BVR_c$ filters. J004331 was observed with a 150 mm Newtonian telescope, also at $f/4$, coupled with an ASI ZWO 183MM Pro CCD and standard $VR_c$ filters. Both setups incorporated a field corrector during observations. Data were acquired using a $4 \times 4$ binning mode. The average CCD operating temperatures were $-10^\circ$C for LZ Leo and $0^\circ$C for J004331. Preprocessing and standard data reduction, including dark, bias, and flat-field corrections, were carried out with Muniwin 2.1.35, Siril\footnote{\url{https://siril.org/}}, and Prism v.10 software.

\vspace{0.3cm}
\subsection{TESS Observations}
NASA launched the Transiting Exoplanet Survey Satellite (TESS) in 2018 to search for exoplanets across the sky \citep{2010AAS...21545006R, 2018AJ....156..102S}. Equipped with four wide-field cameras, TESS observes different regions of the sky, dedicating about 27.4 days to each sector. In this study, we used time-series data from TESS for the binary systems BF Dor, LZ Leo, and J004331. The TESS light curves are provided in the “TESS:T” passband, which covers a broad wavelength range of 600–1000 nm \citep{2015JATIS...1a4003R}. The TESS sectors used in this work are listed in Table \ref{tess}. All TESS data were obtained from the Mikulski Archive for Space Telescopes (MAST)\footnote{\url{https://mast.stsci.edu/portal/Mashup/Clients/Mast/Portal.html}}. We extracted the light curves using the Lightkurve software package\footnote{\url{https://docs.lightkurve.org}} and applied detrending based on the TESS Science Processing Operations Center (SPOC) pipeline \citep{2016SPIE.9913E..3EJ}.

\begin{table*}
\renewcommand\arraystretch{1.2}
\caption{Specifications of the ground-based photometric observations, including the observation dates, filters, exposure times (in seconds), maximum apparent magnitudes in the $V$ band, and observatory sites for each target system.}
\centering
\begin{center}
\footnotesize
\begin{tabular}{c c c c c c}
\hline
System & Observation Date & Filter & Exposure time(s) & $V_{\textit{max}}$(mag) & Observatory\\
\hline
BF Dor	& 2024 Dec. 27 - 2025 Feb. 22 & $BVR_cI_c$ & $B(30)$, $V(15)$, $R_c(10)$, $I_c(12)$ & 13.08(7) & CASLEO\\
J014306	& 2024 Oct. 15  & $BVR_cI_c$ & $B(90)$, $V(50)$, $R_c(35)$, $I_c(30)$ & 14.25(8) & SPM\\
LZ Leo	& 2024 April 4, 11 & $BVR_c$ & $B(120)$, $V(120)$, $R_c(120)$ & 13.20(5) & OABAC\\
J201317 & 2024 Jul. 23, and Jul. 31 & $BVR_cI_c$ & $B(90)$, $V(50)$, $R_c(35)$, $I_c(30)$ & 13.02(10) & SPM\\
J004331 & 2024 Sept. 10, 15, 20, and Oct. 10 & $VR_c$ & $V(180)$, $R_c(180)$ & 13.65(11) & OABAC\\
\hline
\end{tabular}
\end{center}
\label{observations}
\end{table*}

\begin{table*}
\renewcommand\arraystretch{1.2}
\caption{List the comparison and check stars in the ground-based observations.}
\centering
\begin{center}
\footnotesize
\begin{tabular}{c c c c c}
\hline
System & Star type & Star Name & RA$.^\circ$(J2000) & DEC$.^\circ$(J2000)\\
\hline
BF Dor & Comparison	& UCAC4 116-011637 & 91.378580 & -66.844710\\
BF Dor & Check	& 2MASS 06054182-6650129 & 91.424289 & -66.836978\\
J014306 & Comparison & 2MASSJ01433748+3836044 & 25.906183 & 38.601202\\
J014306 & Check	& 2MASSJ01431770+3838529 & 25.823790 & 38.648055\\
LZ Leo  & Comparison & Gaia DR2 615285229435410432 & 149.313674 & 14.246790\\
LZ Leo &	Check	& Gaia DR2 615661679023766528 & 149.287898 & 14.305030\\
J201317  &	Comparison	& 2MASSJ20131800+4642445 & 303.325019 & 46.712383\\
J201317 &	Check	& 2MASSJ20133198+4641317 & 303.383282 & 46.692186\\
J004331  &	Comparison	& Gaia DR2 414729124506961152 & 10.737472 & 50.063308\\
J004331 &	Check	& Gaia DR2 414729468104365056 & 10.850449 & 50.034823\\
\hline
\end{tabular}
\end{center}
\label{stars}
\end{table*}

\begin{table*}
\caption{Specifications of the TESS data used in this study.}
\centering
\begin{center}
\footnotesize
\begin{tabular}{c c c c c c c}
\hline
System & TIC & TESS Sector & Observation Year & Exposure Length & Error Average\\
\hline
BF Dor & 41169654 & 2,5,8,9,12,29,32, & 2018,2018,2019,2019,2019,2020,2020, & 1800,1800,1800,1800,1800,600,600, & 0.0391\\
&& 35,39,62,63,64,65, & 2021,2021,2023,2023,2023,2023, & 600,600,200,200,200,200, &\\
&& 66,67,68,69,89,90 & 2023,2023,2023,2023,2025,2025 & 200,200,200,200,200,200 &\\
LZ Leo & 358669926 & 45,46,72 & 2021,2021,2023 & 600,600,200 & 0.1874\\
J004331 & 240745396 & 17,57,58,84,85 & 2019,2022,2022,2024,2024 & 1800,200,200,200,200 & 0.0519\\
\hline
\end{tabular}
\end{center}
\label{tess}
\end{table*}

\vspace{0.6cm}
\section{Investigating Orbital Period Variations}
Conducting further analysis of these targets required gathering as many eclipse timing measurements as possible from photometric surveys, including both ground-based and space-based observations. Using the available eclipse timings, we analyzed the orbital period variations of all five targets. However, for J201317, only our own observed minima were available, and the limited number of eclipsing times prevented any meaningful further analysis. Moreover, eclipse timing data for LZ Leo were gathered from the VarAstro database\footnote{\url{http://var.astro.cz}}, where no uncertainties are reported for the available measurements. Eclipse timings were directly determined from the TESS 2-minute and 10-minute cadence data. In contrast, for the more sparsely sampled 30-minute cadence data, we first applied the phase-folding approach outlined by \cite{Li_2020} to align the observations within a single orbital cycle. Once aligned, eclipse timings were extracted. The minima times are commonly extracted using the \cite{kw1956} (KW) method, which calculates mid-eclipse times from binary system light curves. Although widely used, the KW method tends to produce underestimated error values (\citealt{2012AN....333..754P}, \citealt{2018MNRAS.480.4557L}). Moreover, its performance can be inadequate for asymmetric or incomplete light curves (\citealt{mikulavsek2013kwee}). To address these issues, we fitted Gaussian and Cauchy distribution models, based on \cite{2021AstL...47..402P} approach, to selected portions of the light curves containing the minima. Uncertainties were estimated via MCMC sampling, implemented in Python using the  emcee package (\citealt{2013PASP..125..306F}).

\begin{figure*}
\centering
\includegraphics[width=0.99\textwidth]{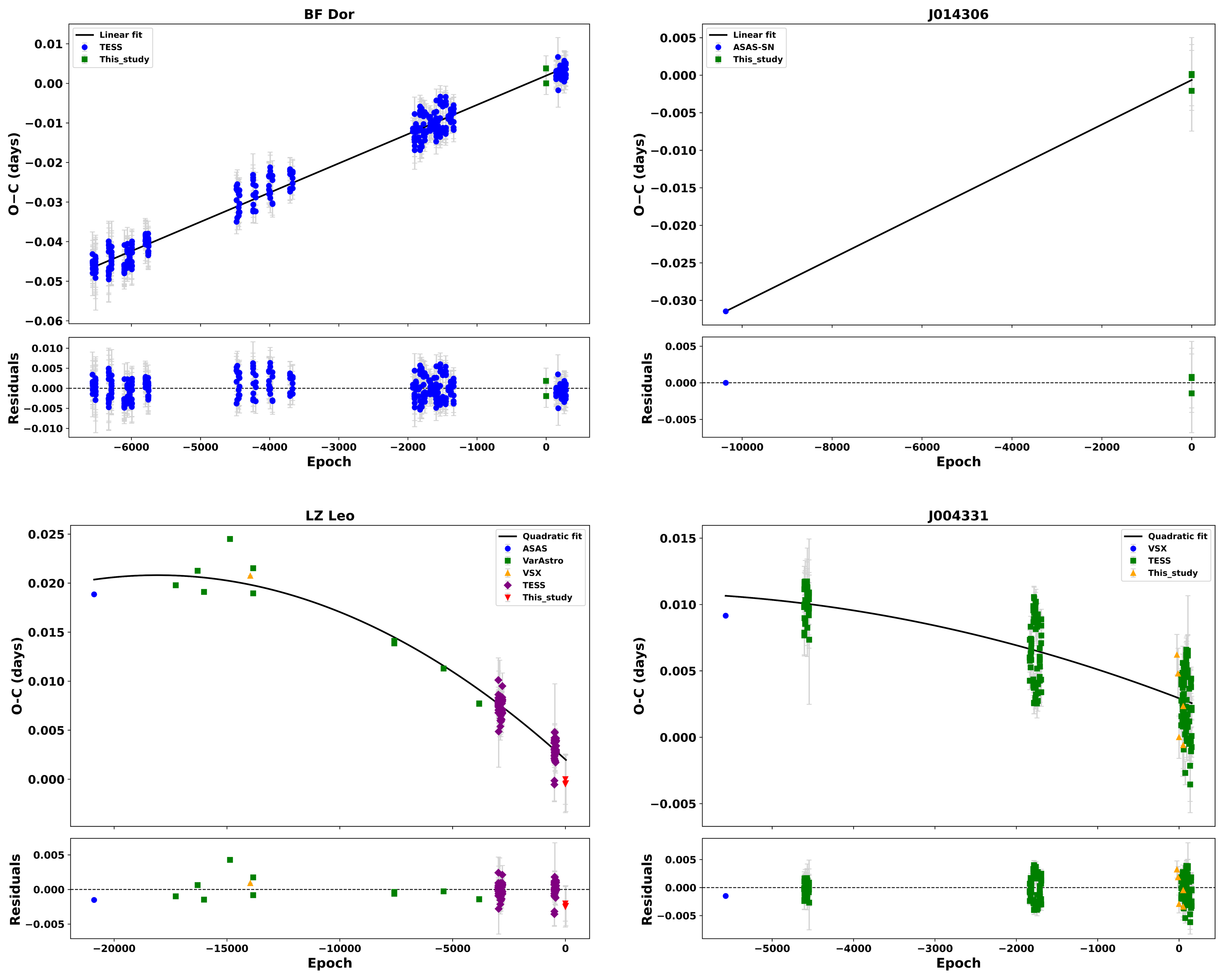}
\caption{The O-C diagrams of the target systems, with residuals at the bottom}.
\label{O-CFigs}
\end{figure*}

To ensure consistency in time measurements across all datasets, we converted Heliocentric Julian Dates ($HJD$) to Barycentric Julian Dates in Barycentric Dynamical Time ($\mathrm{BJD_{TDB}}$) using an online conversion tool\footnote{\url{https://astroutils.astronomy.osu.edu/time/hjd2bjd.html}}. This step was necessary because our data included both time formats. The eclipse timings derived from our observations are presented in Table \ref{min}. A machine-readable version of the extracted and compiled eclipse timing data for the target binary systems is available. Next, we computed the O-C values based on the reference ephemeris:

\begin{equation}\label{eq:OC}
\begin{aligned}
BJD=BJD_0+P\times E,
\end{aligned}
\end{equation}

\noindent where $BJD$ represents the observed eclipse timing, $BJD_0$ (given in the second column of Table \ref{ephemeris}) denotes the initial primary eclipse time, and $P$ (provided in the third column of Table \ref{ephemeris}) corresponds to the orbital period. The calculated epochs and O–C values are provided in Table \ref{min} and are also available in an online machine-readable format. The corresponding O–C diagrams are shown in Figure \ref{O-CFigs}.

We find that two of the targets, BF Dor and J014306, exhibit linear trends, while LZ Leo and J004331 display parabolic variations. For the targets with parabolic behavior, we applied the following equation for the O-C fitting:

\begin{equation}\label{parabolic}
\begin{aligned}
O-C=\Delta{T_0}+\Delta{P_0}\times E+\frac{\beta}{2}{E^2}.
\end{aligned}
\end{equation}

The revised ephemerides are presented in Table \ref{ephemeris}. For the two systems displaying parabolic variations, the corresponding fitted parameters and estimated mass transfer rates are detailed in Table \ref{mass-transfer}. We found that the LZ Leo and J004331 systems exhibit a long-term decrease.

\renewcommand\arraystretch{1.2}
\begin{table*}
\caption{The times of minima extracted from our ground-based observations.}
\centering
\small
\begin{tabular}{c c c c c}
\hline
System & Min.($BJD_{TDB}$) & Error & Epoch & O-C\\ 
\hline
BF Dor	&	2460671.6576	&	0.0032	&	-0.5	&	0.0038	\\
	&	2460671.8299	&	0.0028	&	0	&	0	\\
J014306	&	2460598.7223	&	0.0049	&	-1	&	0.0002	\\
	&	2460598.8597	&	0.0054	&	-0.5	&	-0.0021	\\
	&	2460599.0014	&	0.0041	&	0	&	0	\\
LZ Leo	&	2460405.3129	&	0.0026	&	0	&	0	\\
	&	2460405.4552	&	0.0029	&	0.5	&	-0.0005	\\
	&	2460412.4089	&	0.0029	&	24	&	-0.0004	\\
J201317	&	2460514.8709	&	0.0009	&	0	&	0	\\
	&	2460522.7348	&	0.0006	&	26.5	&	-0.0002	\\
	&	2460522.8843	&	0.0007	&	27	&	0.0009	\\
J004331	&	2460564.3977	&	0.0015	&	-25.5	&	0.0062	\\
	&	2460569.5024	&	0.0019	&	-12.5	&	0.0048	\\
	&	2460574.4073	&	0.0016	&	0	&	0	\\
	&	2460594.4411	&	0.0011	&	51	&	0.0023	\\
	&	2460594.6346	&	0.0018	&	51.5	&	-0.0006	\\
\hline
\end{tabular}
\label{min}
\end{table*}

\renewcommand\arraystretch{1.2}
\begin{table*}
\caption{Reference and new ephemeris of the five systems. The reference times of minimum ($t_0$) were obtained from our observations in this study.}
\centering
\small
\begin{tabular}{c|cc|cc}
\hline
System& \multicolumn{2}{c|}{Reference ephemeris}& \multicolumn{2}{c}{New ephemeris}\\ 
&$t_0(BJD_{TDB})$&Period(day)/Source& Corrected $t_0(BJD_{TDB})$&New Period(day)\\ 
\hline
BF Dor &  2460671.8299(8) & 0.352271/VSX & 2460671.8319(3) &  0.35227839(6)\\
J014306 &  2460599.0014(11) & 0.279231/ASAS-SN & 2460599.0008(5) & 0.27923397(10)\\
LZ Leo &  2460405.3129(26) & 0.295682/VSX & 2460405.3149(2) & 0.29567992(6)\\
J201317 &  2460514.8709(9) & 0.2967596/VSX & - & -\\
J004331 &  2460574.4073(16) & 0.3927752/VSX & 2460574.4102(2) & 0.3927729(3)\\
\hline
\end{tabular}
\label{ephemeris}
\end{table*}

\renewcommand\arraystretch{1.2}
\begin{table*}
\caption{The O-C fitting coefficients and mass transfer rate.}
\centering
\small
\begin{tabular}{ccccccccc}
\hline
Parameter& $\Delta{T_0}$& Error& $\Delta{P_0}$& Error& $\beta$&Error & $dM_1/dt$&Error\\ 
&$(\times {10^{-4}} d)$&& $(\times {10^{-7}} d)$&& $(\times {10^{-7}} d$ $ {yr^{-1}})$& & $(\times {10^{-7}} M_\odot$ $ {yr^{-1}})$&\\ 
\hline
LZ Leo & 20.1 & 1.4 & -20.8 & 0.6 & -1.4 & 0.1 & 2.2 & 3.1\\
J004331 & 29.4 & 2.3 & -23.4 & 2.9 & -3.2 & 1.2 & -0.4 & 0.4\\
\hline
\end{tabular}
\label{mass-transfer}
\end{table*}

\vspace{0.6cm}
\section{Light Curve Solution}
The BSN application version 1.0 (\citealt{paki2025bsn}), specifically developed for modeling contact binary stars, was used to analyze the photometric light curves of the target systems. This application offers an expanded feature set and a more intuitive interface while adhering to established scientific standards. Currently, it only supports the Windows operating system.

For the target systems, the contact configuration was selected, as the observed light curve shapes, catalog classifications, and their short orbital periods all indicate that these systems are in physical and thermal contact. Time was converted to phase for the light curves using ephemerides in Table \ref{ephemeris}. The gravity-darkening coefficient was fixed at $g_1 = g_2 = 0.32$ following \cite{1967ZA.....65...89L}, and the bolometric albedo was set to $A_1 = A_2 = 0.5$ as per \cite{1969AcA....19..245R}. The stellar atmosphere model adopted was that of \cite{2004AA...419..725C}. In the BSN application, linear and logarithmic limb-darkening formulations are implemented using coefficients adopted from the tabulations provided by \cite{1993AJ....106.2096V}.

For the analysis, the initial effective temperatures ($T$) were obtained from the Gaia DR3 database (Table \ref{systemsinfo}). The LZ Leo system lacked a temperature entry in Gaia DR3; for this case, the temperature of 5113(305) K from version 8.2 of the TESS Input Catalog (TIC) was used instead. For LZ Leo, an independent spectroscopic estimate from the LAMOST-Low Resolution Spectroscopy(LRS) survey (5145 $\pm$ 38 K) is also available, which is in good agreement with the TIC value, supporting the reliability of the adopted temperature. It was assumed that the temperatures reported in Gaia DR3 and TIC correspond to the hotter components of each system, as indicated by the depths of the light curve minima. The effective temperature of the cooler star was estimated based on the difference in depth between the primary and secondary minima in the light curves.

\begin{figure*}
\centering
\includegraphics[width=0.99\textwidth]{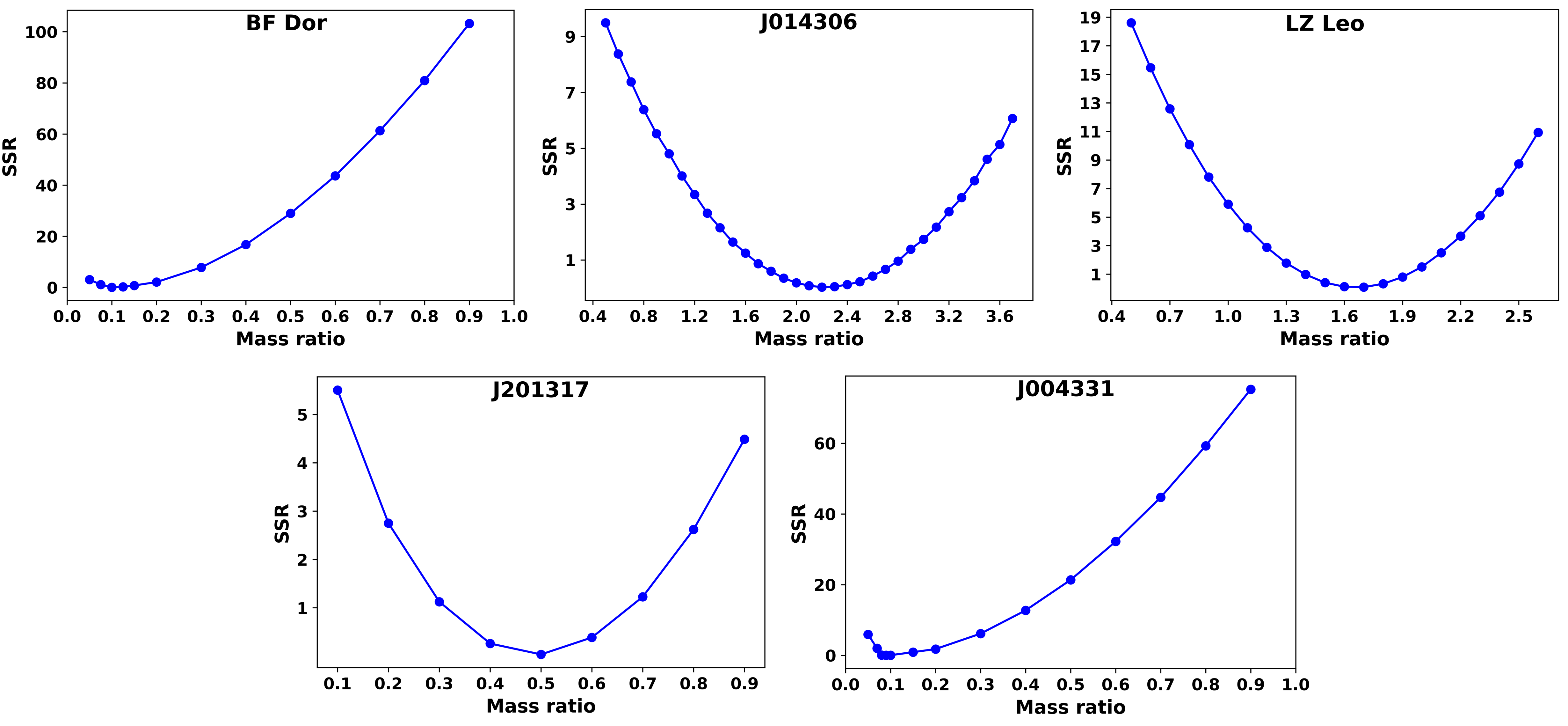}
\caption{The sum of squared residuals as a function of mass ratio.}
\label{q-diagrams}
\end{figure*}

For each system, the mass ratio ($q$)-search method was employed to estimate the initial mass ratio and its possible range, which served as the starting point for the light curve analysis (\citealt{2005ApSS.296..221T}). A wide range of $q$ values, from 0.05 to 20, was examined for all target systems. Subsequently, a narrower range was explored to refine the estimate by minimizing the sum of squared residuals between the observed and synthetic light curves. As illustrated in Figure \ref{q-diagrams}, each $q$-search curve shows a clear minimum in the residual sum, indicating the preliminary mass ratio. This initial estimate serves as the starting point for further analysis, with the final mass ratio obtained through iterative light curve modeling techniques. It is worth noting that, according to \cite{2024AJ....168..272P}, photometric analyses of systems with different orbital inclinations can still achieve reliable $q$-search accuracy when iterative methods such as MCMC are employed. Similar investigations have also been presented by \cite{2023MNRAS.519.5760L} and \cite{2021AJ....162...13L} in this regard.

An asymmetry between the light curve maxima, known as the O'Connell effect, was observed only in the J014306 system. To achieve an acceptable theoretical fit for this system, a cool starspot on the primary component was required (Table \ref{lc-analysis}). This effect is commonly attributed to magnetic activity on the stellar surface, which leads to the formation of starspots (\citealt{1951PRCO....2...85O}). Although this explanation is widely accepted, other physical interpretations have also been proposed to explain the phenomenon more comprehensively, such as those suggested by \cite{1990ApJ...355..271Z} and \cite{2003ChJAA...3..142L}.

We used the photometric multiband data and initial parameter values to achieve a satisfactory theoretical fit. The optimization tool in the BSN application was subsequently employed to improve the light curve solution, providing better-constrained parameter estimates compared to the initial values, including the effective temperatures, mass ratio, fillout factor, and orbital inclination. In the single case where a starspot was included in the model, the optimization process also provided a refined determination of its position.

To derive the final solutions and estimate parameter uncertainties, we employed the Markov Chain Monte Carlo (MCMC) method. The BSN application offers substantially higher computational performance in MCMC fitting, generating synthetic light curves over 40 times faster than PHOEBE Python code version 2 (\citealt{paki2025bsn}). This improvement is primarily due to BSN's optimized architecture and the integration of modern computational libraries, while the fundamental approaches for light curve analysis remain consistent with those used in other established binary star modeling tools. In the MCMC simulations, we used 24 walkers and 2000 iterations to sample five main parameters ($T_{1,2}$, $q$, $f$, and $i$), providing both their estimated values and corresponding uncertainties.

\begin{figure*}
\centering
\includegraphics[width=0.41\textwidth]{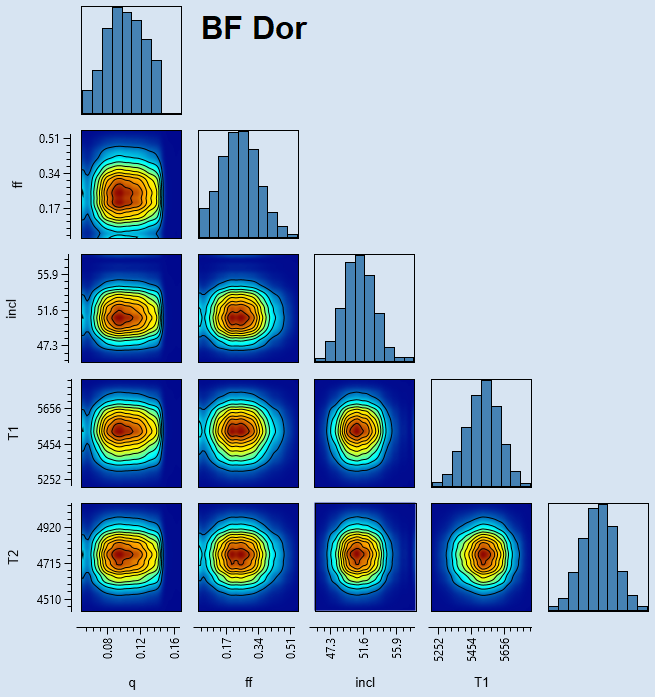}
\includegraphics[width=0.41\textwidth]{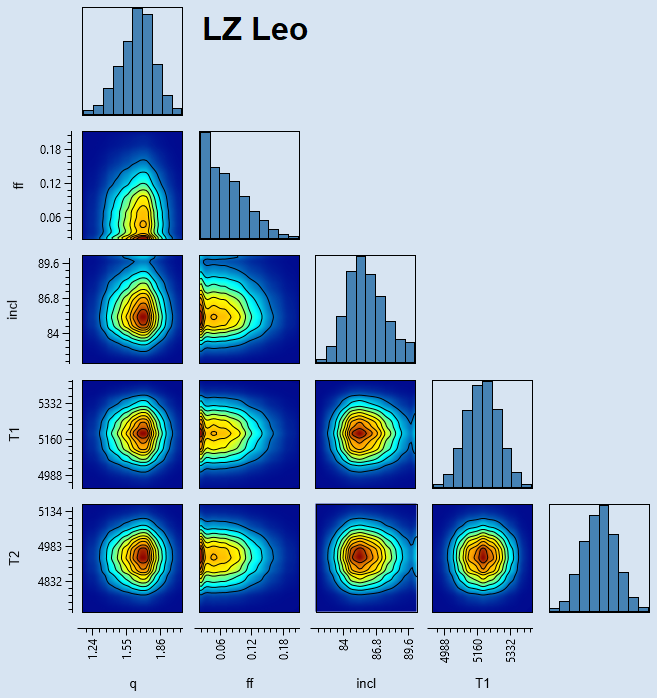}
\includegraphics[width=0.41\textwidth]{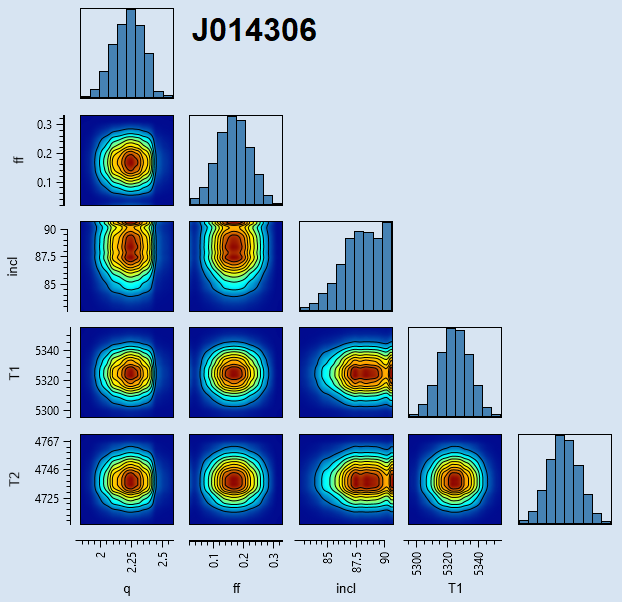}
\includegraphics[width=0.41\textwidth]{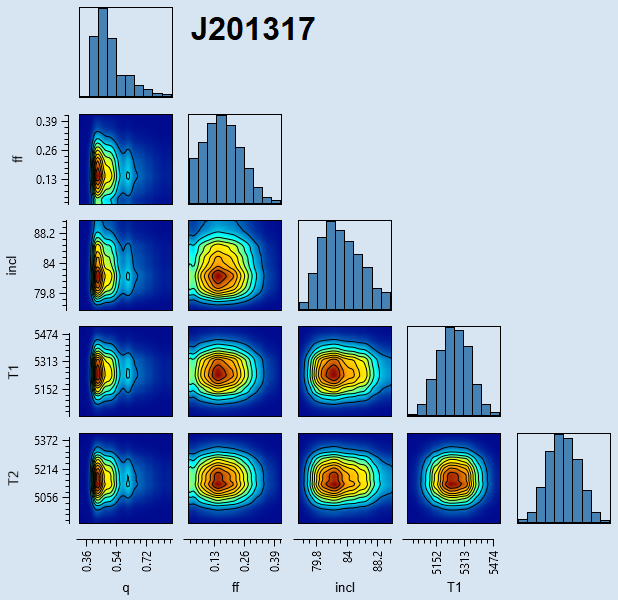}
\includegraphics[width=0.415\textwidth]{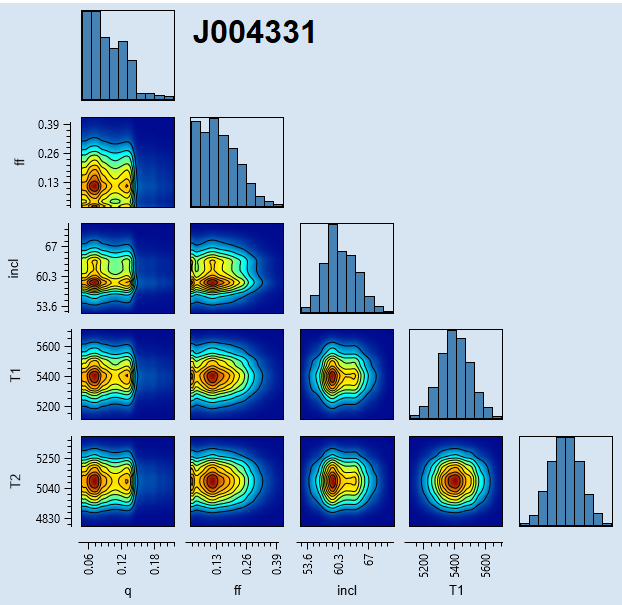}
\caption{The corner plots of the target systems were determined by MCMC modeling.}
\label{corner}
\end{figure*}

\begin{figure*}
\centering
\includegraphics[width=0.49\textwidth]{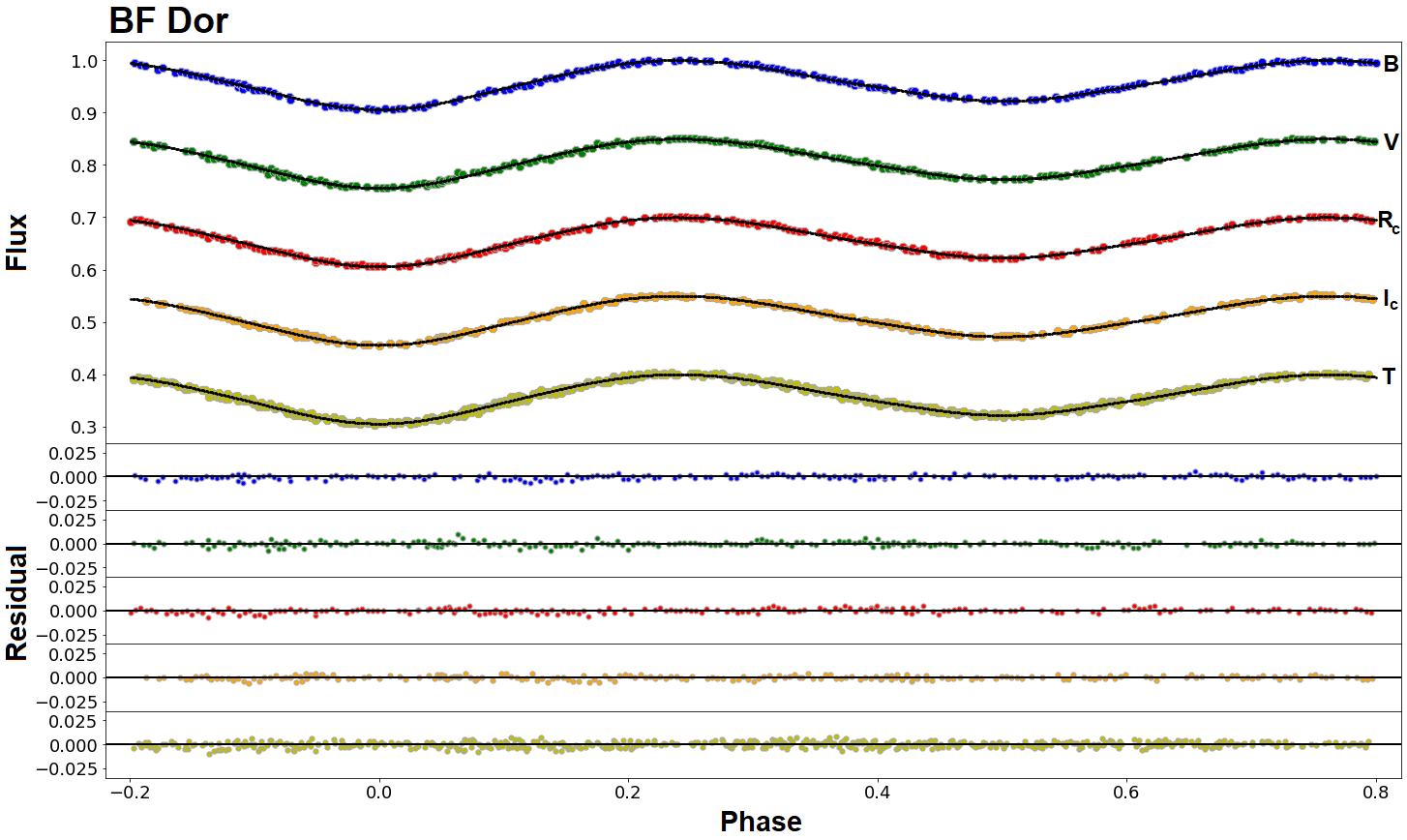}
\includegraphics[width=0.49\textwidth]{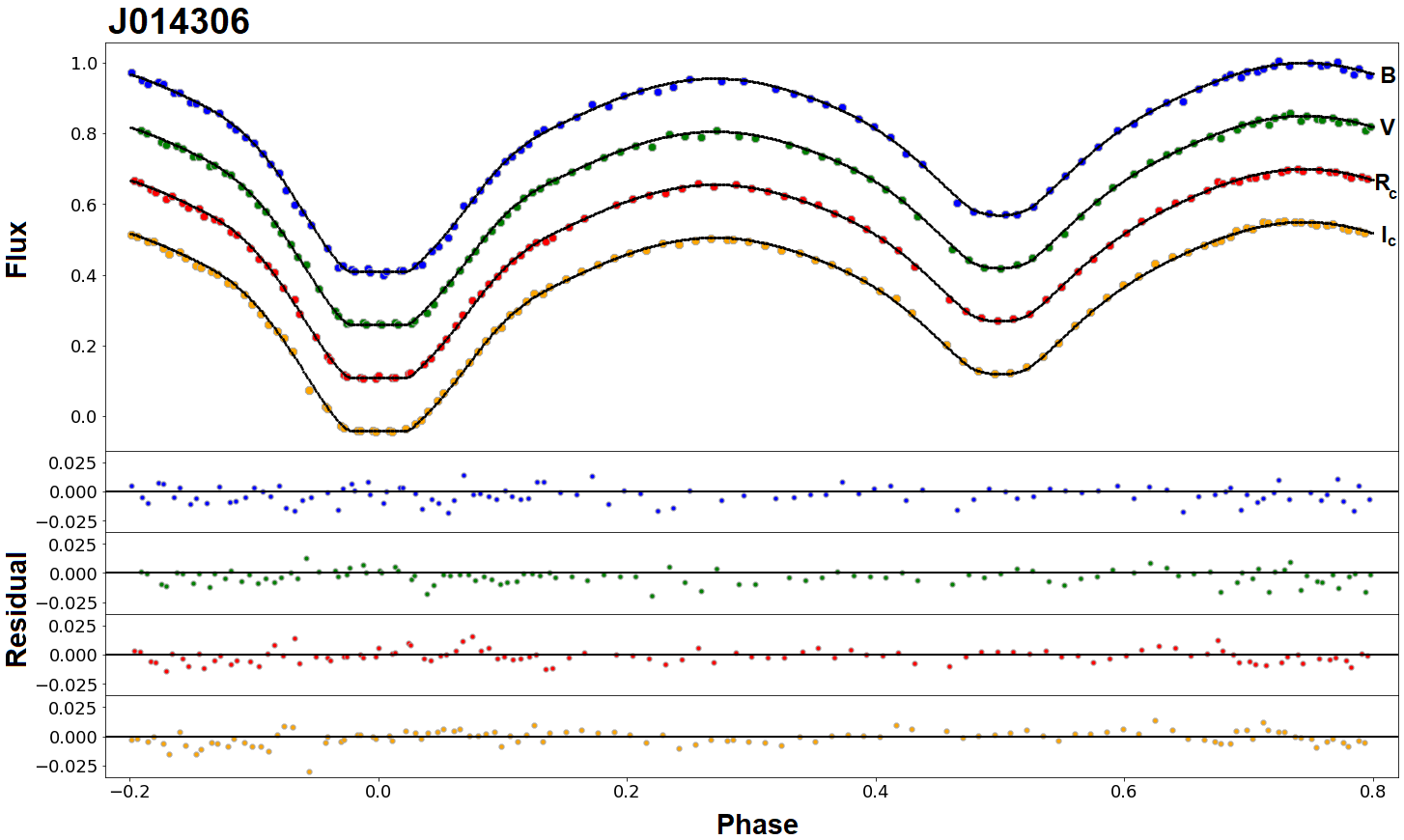}
\includegraphics[width=0.49\textwidth]{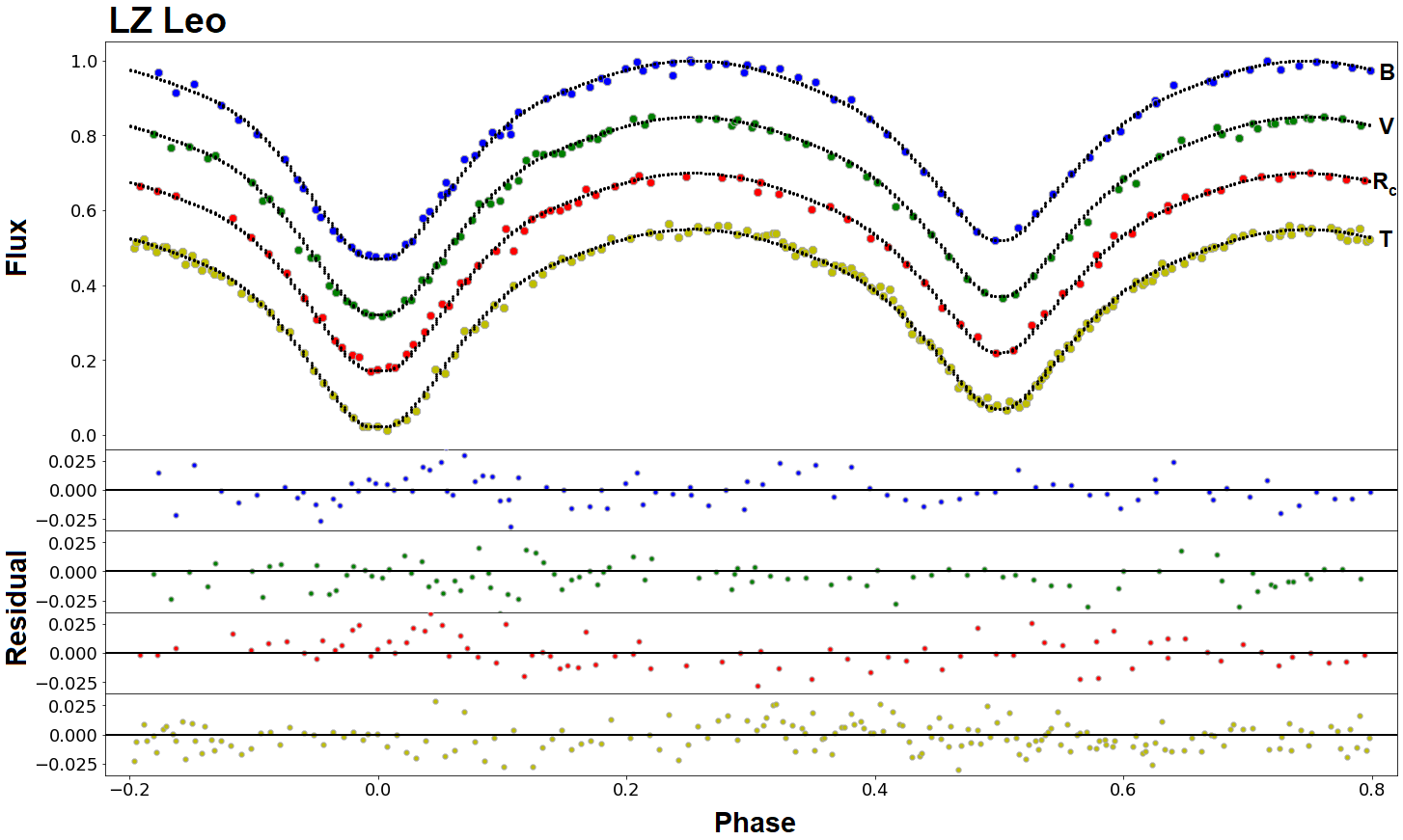}
\includegraphics[width=0.49\textwidth]{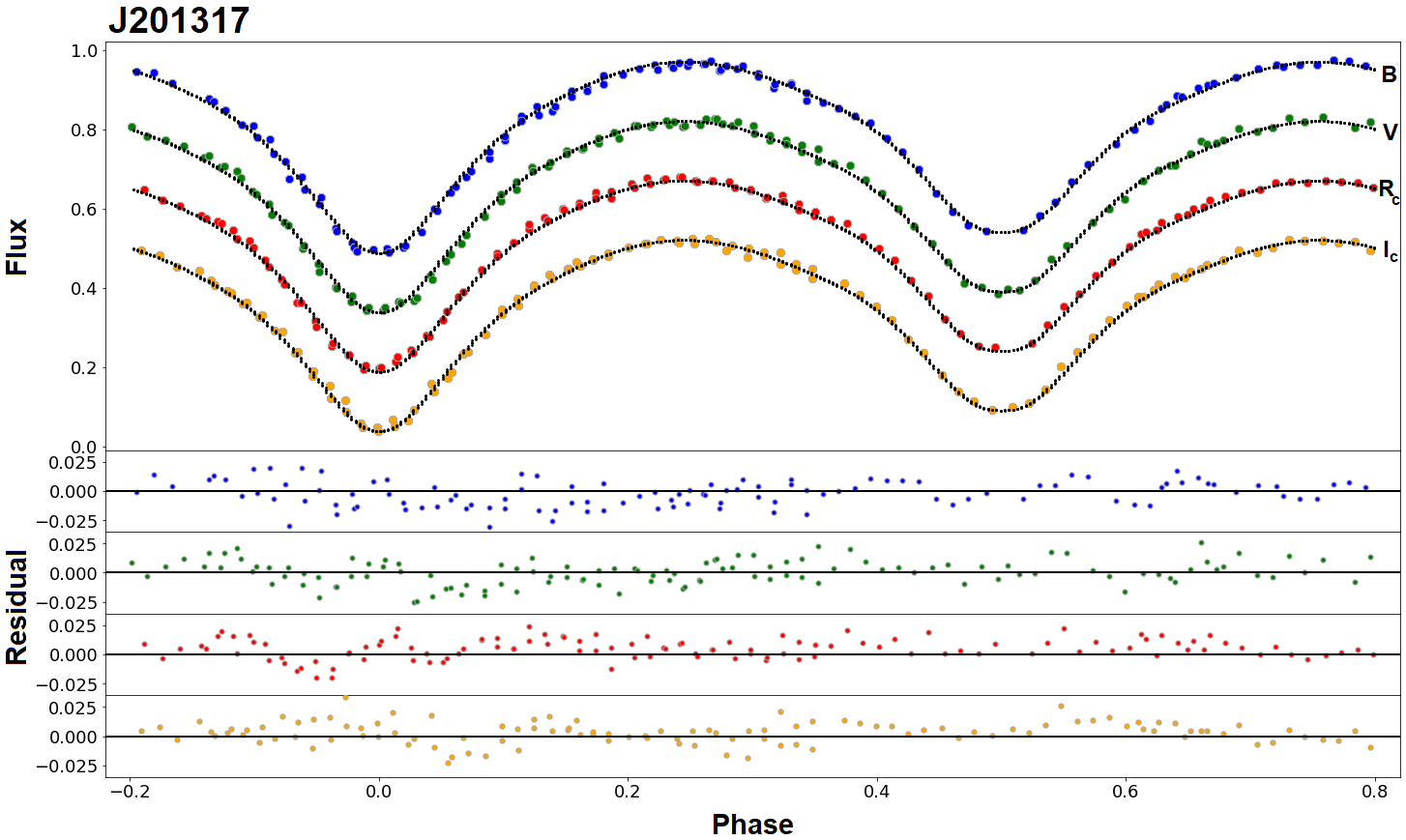}
\includegraphics[width=0.49\textwidth]{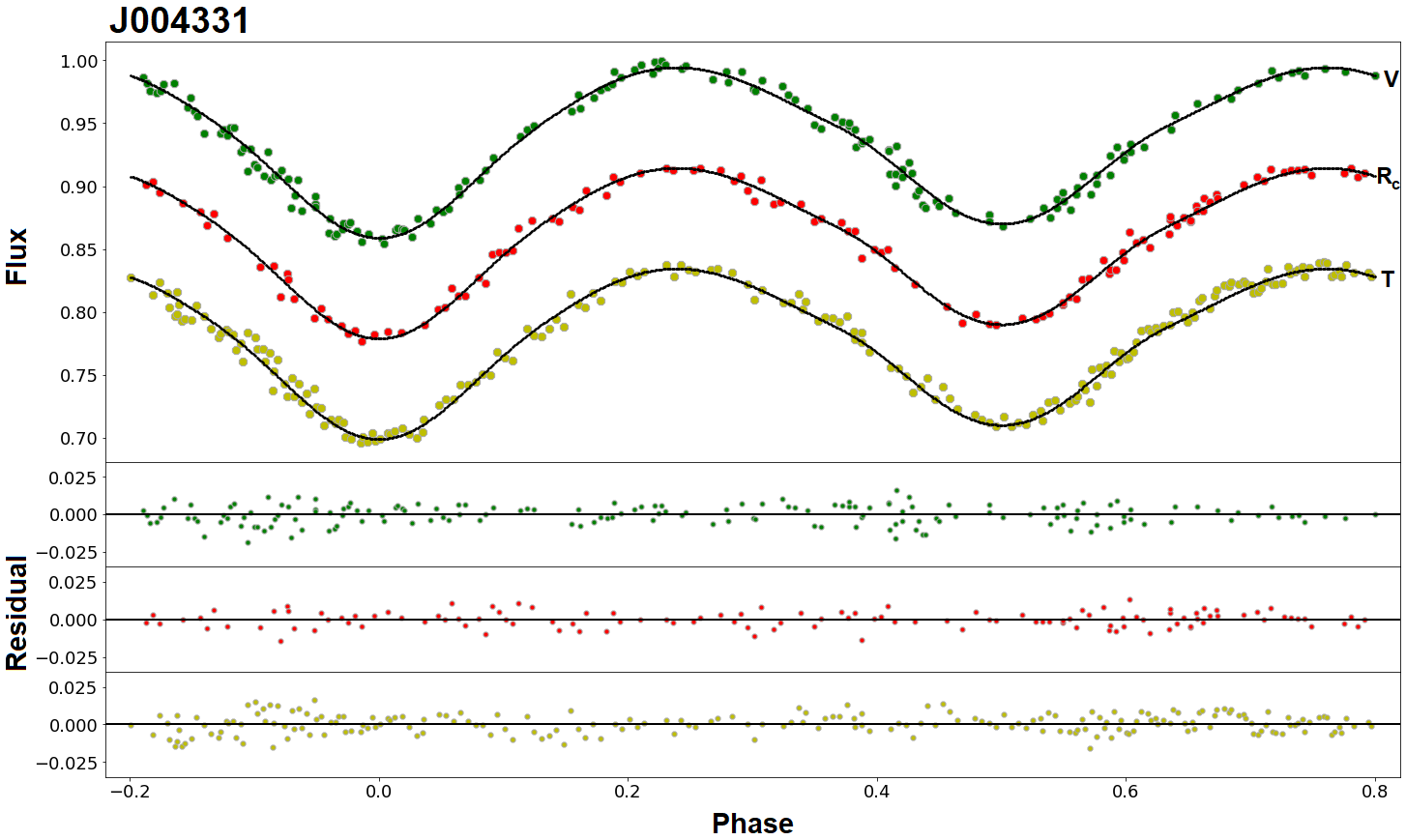}
\caption{The colored dots represent the observed light curves of the systems in different filters, and the synthetic light curves, generated using the light curve solutions, are also shown. Residuals are shown at the bottom of each panel.}
\label{LCs}
\end{figure*}

\begin{figure*}
\centering
\includegraphics[width=0.82\textwidth]{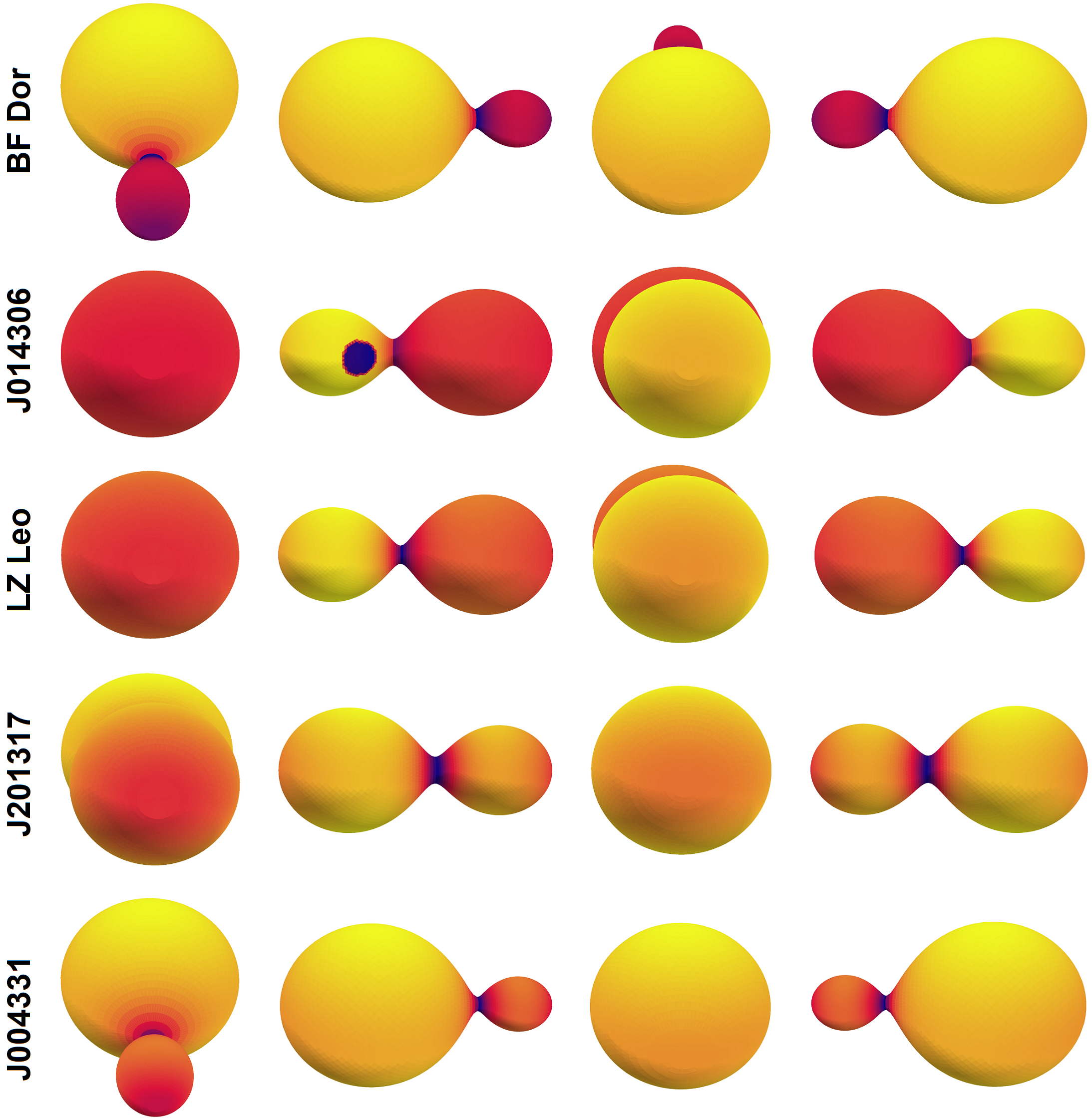}
\caption{Three-dimensional views of the stars in the target binary systems at orbital phases 0.0, 0.25, 0.5, and 0.75, respectively.}
\label{3d}
\end{figure*}

The corner plots for the target systems are shown in Figure \ref{corner}, providing a visualization of the parameter distributions and correlations from the MCMC analysis. Table \ref{lc-analysis} presented the results of the light curve modeling, including the estimated parameters and their uncertainties. Figure \ref{LCs} illustrates the final synthetic light curves overlaid on the observed data for the binary systems. Additionally, three-dimensional (3D) visualizations of the binary systems are presented in Figure \ref{3d}.

\begin{table*}
\renewcommand\arraystretch{1.8}
\caption{Light curve solutions of the target binary stars.}
\centering
\begin{center}
\footnotesize
\begin{tabular}{c c c c c c}
\hline
Parameter & BF Dor & J014306 & LZ Leo & J201317 & J004331\\
\hline
$T_{1}$ (K)	&	$5534_{\rm-(107)}^{+(100)}$	&	$5325_{\rm-(17)}^{+(17)}$	&	$5187_{\rm-(87)}^{+(86)}$	&	$5234_{\rm-(81)}^{+(82)}$	&	$5401_{\rm-(97)}^{+(104)}$	\\
$T_{2}$ (K)	&	$4754_{\rm-(105)}^{+(101)}$	&	$4736_{\rm-(16)}^{+(17)}$	&	$4929_{\rm-(77)}^{+(75)}$	&	$5148_{\rm-(77)}^{+(81)}$	&	$5091_{\rm-(105)}^{+(106)}$	\\
$q=M_2/M_1$	&	$0.100_{\rm-(23)}^{+(24)}$	&	$2.232_{\rm-(134)}^{+(120)}$	&	$1.665_{\rm-(173)}^{+(134)}$	& $0.487_{\rm-(56)}^{+(130)}$	&	$0.093_{\rm-(29)}^{+(41)}$	\\
$i^{\circ}$	&	$50.94_{\rm-(1.92)}^{+(1.98)}$	&	$87.71_{\rm-(1.74)}^{+(1.66)}$	&	$85.69_{\rm-(1.41)}^{+(1.89)}$	&	$83.01_{\rm-(2.45)}^{+(3.27)}$	&	$60.54_{\rm-(2.82)}^{+(3.91)}$	\\
$f$	&	$0.229_{\rm-(98)}^{+(100)}$	&	$0.169_{\rm-(54)}^{+(55)}$	&	$0.066_{\rm-(38)}^{+(48)}$	&	$0.161_{\rm-(79)}^{+(86)}$	&	$0.126_{\rm-(73)}^{+(92)}$	\\
$\Omega_1=\Omega_2$	&	1.945(33)	&	5.475(302)	&	4.731(293)	&	2.803(146)	&	1.930(10)	\\
$l_1/l_{tot}$($V$)	&	0.935(11)	&	0.458(5)	&	0.445(7)	&	0.675(9)	&	0.918(7)	\\
$l_2/l_{tot}$($V$) 	&	0.065(3)	&	0.542(5)	&	0.555(7)	&	0.325(7)	&	0.082(2)	\\
$r_{(mean)1}$	&	0.588(7)	&	0.322(12)	&	0.339(5)	&	0.454(11)	&	0.591(3)	\\
$r_{(mean)2}$	&	0.213(8)	&	0.462(11)	&	0.429(6)	&	0.329(13)	&	0.205(4)	\\
\hline													
Component	&	-	&	Primary	&	-	&	-	&	-	\\
$Col.^\circ$(spot)	&	-	&	96(2)	&	-	&	-	&	-	\\
$Long.^\circ$(spot)	&	-	&	312(3)	&	-	&	-	&	-	\\
$Radius^\circ$(spot)	&	-	&	20(1)	&	-	&	-	&	-	\\
$T_{spot}/T_{star}$	&	-	&	0.73(1)	&	-	&	-	&	-	\\
\hline
\end{tabular}
\end{center}
\label{lc-analysis}
\end{table*}

\vspace{0.6cm}
\section{Absolute Parameters}
\subsection{Method}
Accurate determination of absolute parameters plays an important role in studying the evolution of contact systems and exploring the correlations among their physical properties. Several approaches are employed to derive the absolute parameters of contact binaries. One such method, applicable when only photometric data are available, relies on Gaia DR3 parallaxes, as comprehensively discussed by \cite{2024NewA..11002227P}. This technique depends significantly on two factors: the interstellar extinction $A_V$ and the maximum visual magnitude $V_{max}$. The precision of $V_{max}$ is linked to the quality of observational procedures, whereas high values of $A_V$ can hinder achieving reliable accuracy in absolute parameter estimation using Gaia DR3 parallaxes. In this study, we derived $A_V$ from the three-dimensional dust map corresponding to Gaia distances (\citealt{2019ApJ...887...93G}). Consequently, for J201317 system, the $A_V$ values (Table \ref{absolute}) is too large to allow the use of Gaia DR3 parallaxes for reliable parameter estimation. As we aimed to apply a consistent method for target systems, we therefore needed to use a different way.

Another way to estimate absolute parameters from photometric data in contact binary systems is to use empirical parameter relationships. Various studies have proposed such relationships, particularly those describing the connection between orbital period and mass ($P$–$M$), as well as between orbital period and semi-major axis ($P$–$a$). However, the $P$–$M$ relationships are usually reliable only for the more massive primary component, while significant scatter is observed in the corresponding plots for the secondary components (\citealt{2022MNRAS.510.5315P}). However, it is also essential to perform a statistical analysis of the scatter and the strength of the correlations between empirical parameters in order to identify the most suitable empirical parameter relationship for estimating absolute parameters.

The empirical relationships between parameters in contact binary systems remain subject to debate, partly due to diverse analysis techniques and limited sample sizes (e.g., \citealt{2022MNRAS.510.5315P}, \citealt{2024RAA....24a5002P}). In this study, we focus on the relationships between the orbital period ($P$) and key physical parameters, including mass ($M$), radius ($R$), and luminosity ($L$), for each component. Using the sample presented by \citet{2025MNRAS.538.1427P}, which initially included 818 systems, we selected only those with available data for $P$, $M$, $R$, and $L$, resulting in a final sample of 483 systems for analysis. Our goal is to identify the strongest statistically significant relationships to select the most reliable ones for estimating absolute parameters.

\begin{figure*}
\centering
\includegraphics[width=0.99\textwidth]{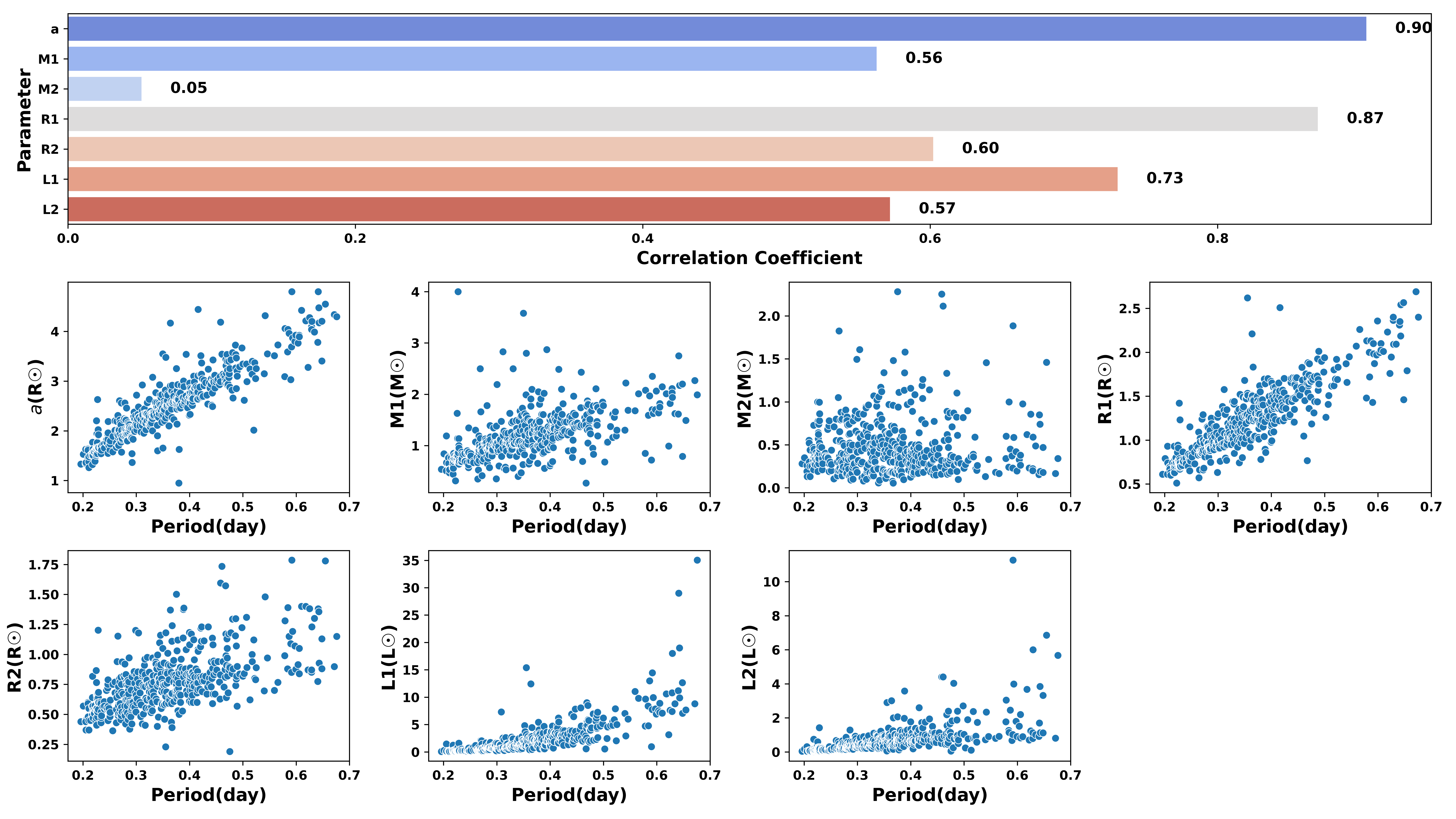}
\caption{Correlation between orbital period and selected physical parameters of contact binaries. The top panel shows the correlation coefficients as a horizontal bar plot, while the bottom panel presents the corresponding scatter plots.}
\label{correlation}
\end{figure*}

The correlation coefficients between the orbital period and selected parameters were computed and presented as a horizontal bar plot (Figure \ref{correlation}, top panel). As the histogram illustrates only positive correlation values, negative correlations are not included in this analysis. The length of each bar represents the strength of the positive linear relationship between the period and the respective parameter. To further investigate these associations, scatter plots of the period versus each parameter are provided in the bottom panel of the same figure (Figure \ref{correlation}, bottom panel). Together, these visualizations offer a comprehensive overview of the magnitude and nature of the dependencies within the dataset.

Based on the results, the relationship between the orbital period and the semi-major axis shows the strongest correlation coefficient of 0.90. These numerical values are also indicated on the histogram bars for clarity.

\subsection{Parameter Estimation}
A revised empirical relationship between $P$ and $a$ for contact binary systems was presented by \cite{2024PASP..136b4201P}. This calibration was derived using observations of 414 systems compiled in the work of \cite{2021ApJS..254...10L}, all exhibiting periods shorter than 0.7 days. The relation is expressed as follows (Equation \ref{p-a}):

\begin{equation}\label{p-a}
a=(0.372_{\rm-0.114}^{+0.113})+(5.914_{\rm-0.298}^{+0.272})\times P,
\end{equation}

\noindent where $a$ is measured in solar radii ($R_\odot$) and $P$ in days.

Since the $P$–$a$ relationship exhibits a stronger correlation compared to the relationships between the orbital period and other parameters such as mass, radius, and luminosity, we adopted this empirical relation for estimating the absolute parameters. The procedure began with Equation \ref{p-a}, in which the orbital period of each target system, was used to calculate the semi-major axis. The stellar masses were then determined using the mass ratio derived from the light curve solutions in combination with Kepler’s third law, as expressed by Equations \ref{eq:M1} and \ref{eq:M2}.

\begin{eqnarray}
M{_1}=\frac{4\pi^2a^3}{GP^2(1+q)}\label{eq:M1},\\
M{_2}=q\times{M{_1}}\label{eq:M2}.
\end{eqnarray}

The radii ($R$) of the stellar components were calculated using the mean fractional radii ($r_{\text{mean, 1}}$ and $r_{\text{mean, 2}}$) provided in Table \ref{lc-analysis}, applying the relation $R_{1,2} = a \times r_{mean1,2}$. With the effective temperature and radius determined, the stellar luminosities were subsequently computed. The absolute bolometric magnitude ($M_{\text{bol}}$) for each component was then derived from the luminosity by employing the standard relation between these parameters, as given in Equation \ref{eq:Mbol}:

\begin{equation}\label{eq:Mbol}
M_{\text{bol, 1,2}} = M_{\text{bol}, \odot} - 2.5 \times \log \left(\frac{L_{1,2}}{L_\odot}\right).
\end{equation}

In Equation \ref{eq:Mbol}, the absolute bolometric magnitude of the Sun is taken as $4.73^{\mathrm{mag}}$, following the value reported by \cite{2010AJ....140.1158T}. The surface gravity ($g$) of each star was computed on a logarithmic scale using the determined masses and radii. Furthermore, the orbital angular momentum ($J_0$) was derived based on the total mass, mass ratio, and orbital period of the systems, according to Equation \ref{eqJ0} as described by \cite{2006MNRAS.373.1483E}:

\begin{equation}\label{eqJ0}
J_0 = \frac{q}{(1+q)^2} \sqrt[3]{\frac{G^2}{2\pi} M^5 P}.
\end{equation}

The absolute parameters obtained for the five contact binary systems are listed in Table \ref{absolute}.

\begin{table*}
\renewcommand\arraystretch{1.8}
\caption{Estimated absolute parameters of the systems.}
\centering
\begin{center}
\footnotesize
\begin{tabular}{c c c c c c}
\hline
Parameter & BF Dor & J014306 & LZ Leo & J201317 & J004331\\
\hline
$M_1(M_\odot)$ &	1.456(415)	&	0.441(139)	&	0.550(169)	&	0.987(302)	&	1.558(425)	\\
$M_2(M_\odot)$ &	0.146(86)	&	0.985(383)	&	0.915(391)	&	0.480(267)	&	0.145(109)	\\
$R_1(R_\odot)$ &	1.444(144)	&	0.652(89)	&	0.719(79)	&	0.966(116)	&	1.593(142)	\\
$R_2(R_\odot)$ &	0.523(67)	&	0.935(114)	&	0.910(99)	&	0.700(95)	&	0.552(58)	\\
$L_1(L_\odot)$ &	1.762(535)	&	0.308(95)	&	0.337(106)	&	0.631(211)	&	1.946(540)	\\
$L_2(L_\odot)$ &	0.126(49)	&	0.396(109)	&	0.440(135)	&	0.310(115)	&	0.185(60)	\\
$M_{bol1}(mag.)$ &	4.115(288)	&	6.010(291)	&	5.910(298)	&	5.230(313)	&	4.007(266)	\\
$M_{bol2}(mag.)$ &	6.980(355)	&	5.735(265)	&	5.620(290)	&	6.001(344)	&	6.563(306)	\\
$log\textit{(g)}_1(cgs)$ &	4.282(26)	&	4.455(8)	&	4.464(26)	&	4.462(18)	&	4.226(30)	\\
$log\textit{(g)}_2(cgs)$ &	4.164(98)	&	4.490(43)	&	4.481(65)	&	4.430(81)	&	4.114(157)	\\
$a(R_\odot)$ &	2.455(214)	&	2.023(193)	&	2.121(198)	&	2.127(198)	&	2.695(225)	\\
$J_0(\text{g.cm}^2/\text{s})$ &	51.200(272)	&	51.495(216)	&	51.563(224)	&	51.538(260)	&	51.235(309)	\\
\hline
$A_V(mag.)$ & 0.187(1) &	0.130(1) &	0.062(1) &	1.323(18) &	0.364(3)\\
\hline
\end{tabular}
\end{center}
\label{absolute}
\end{table*}

\vspace{0.6cm}
\section{Discussion and Conclusion}
In this work, we present the first analysis of photometric light curves, investigate orbital period variations, and estimate the absolute parameters of five target contact binary stars located in the northern and southern hemisphere. The outcomes of these analyses form the basis for the following discussion and conclusions:

A) Based on the analysis of the O–C diagrams, two of the targets exhibit linear variations consistent with the available minima (BF Dor and J014306), while the other two show parabolic trends (LZ Leo and J004331), both indicating a long-term decrease in their orbital periods.

To analyze the orbital period variations of LZ Leo and J004331 targets, we assumed that the observed orbital period variations are driven by conservative mass transfer. To estimate the mass transfer rates, we employed Equation \ref{eqPM} \citep{k1958}:

\begin{equation}\label{eqPM}
\frac{\dot{P}}{P} = -3 \dot{M} \left( \frac{1}{M_1} - \frac{1}{M_2} \right),
\end{equation}

\noindent where the results of these calculations are summarized in Table \ref{mass-transfer}. Given the relatively short time span covered by the O–C diagrams for all targets, additional observations will be necessary to verify these findings in the future.

B) Light curve analysis was performed using the BSN application and the MCMC algorithm. Asymmetry in the maxima is a common feature in contact binary systems, often associated with surface phenomena such as starspots. Among the five systems studied, only one exhibited significant asymmetry that required the inclusion of a cold starspot in the model. This type of asymmetry is commonly referred to as the O'Connell effect, which may be explained by stellar surface inhomogeneities such as starspots (\citealt{1951PRCO....2...85O}).

Based on the light curve solutions, the stellar effective temperatures of the target systems span from 4736 K to 5534 K. Among them, the systems J201317 and BF Dor exhibit the smallest and largest temperature differences ($\Delta T=|T_1-T2|$) between their components, respectively (Table \ref{conclusion-tab}).

\begin{figure}
\centering
\includegraphics[width=0.5\textwidth]{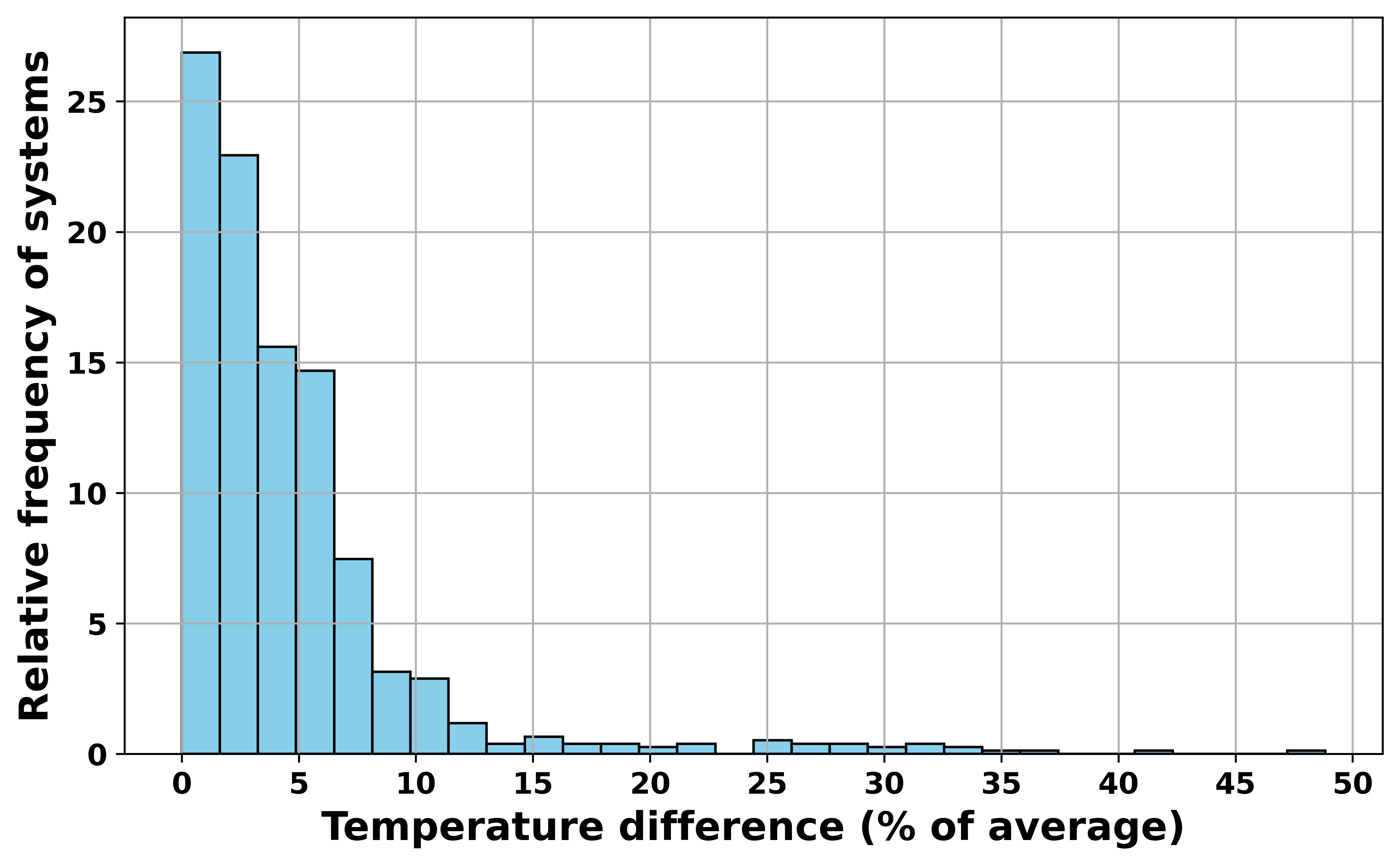}
\caption{Histogram showing the relative frequency distribution of temperature difference percentages among contact binary systems. The x-axis represents the temperature difference as a percentage of the average component temperature, while the y-axis shows the fraction of systems in each bin.}
\label{histogram}
\end{figure}

It is well established that in contact binary systems, the stellar components are expected to have close surface temperatures due to ongoing mass and energy exchange through the common envelope. Nevertheless, a statistical view of the temperature differences between the components of contact binary systems can be informative and insightful. For this purpose, we prepared a sample of 763 contact binary systems characterized by orbital periods shorter than 0.7 days and hotter components with effective temperatures below 8000 K from the \cite{2025MNRAS.538.1427P} study.
The temperature difference percentage (DT\%) quantifies the relative difference in surface temperature between the two stellar components in contact binary systems. The median DT\% in the sample is approximately 3.29\%, indicating that half of the systems exhibit relatively small temperature contrasts. Notably, about 90\% of the systems have a DT\% below 9.38\%, demonstrating that the vast majority of contact binaries maintain close thermal equilibrium. Furthermore, only a small fraction of systems show large temperature differences, with the maximum DT\% reaching nearly 49\%. These results suggest that most contact binaries currently tend to have components with close or equal temperatures, although a minority exhibit significant thermal disparities. The detailed results are presented in Table \ref{dt_stats} and Figure \ref{histogram}. The DT\% of the five targey systems — BF Dor (15.16\%), J014306 (11.71\%), LZ Leo (5.10\%), J201317 (1.66\%), and J004331 (5.91\%) — span a range from close thermal equilibrium to moderately higher contrasts. Most of these systems fall within the 90th percentile of the overall sample, while BF Dor and J014306 exhibit temperature differences approaching the upper range of observed values. The relatively large effective temperature differences observed in BF Dor and J014306 may arise from different physical causes. In the case of BF Dor, the extreme mass ratio of $q=0.1$ likely leads to inefficient energy transfer between the components, preventing them from reaching full thermal equilibrium (\citealt{2023A&A...672A.175F}). For J014306, however, the temperature difference may instead reflect variations in the evolutionary status of the components or other system-specific factors that affect the efficiency of energy redistribution. Both systems warrant further studies to better understand the origin of these surface temperature differences.

The spectral classifications of the component stars were determined using the temperature scales provided by \cite{2000asqu.book.....C} and \cite{2018MNRAS.479.5491E}, as listed in Table \ref{conclusion-tab}.

\begin{table*}
\renewcommand\arraystretch{1.2}
\caption{Conclusions derived for each target system.}
\centering
\begin{center}
\footnotesize
\begin{tabular}{c c c c c c}
\hline
Parameter & BF Dor & J014306 & LZ Leo & J201317 & J004331\\
\hline
$\Delta T=|T_1-T_2|$ ($K$) & 780 & 589 & 258 & 86 & 310\\
Spectral category & G8-K3 & K0-K3 & K0-K2 & K0-K1 & G8-K1\\
Subtype & A & W & W & A & A\\
$M_{1i}$ ($M_{\odot}$) & 1.03 & 0.71 & 0.74 & 0.75 & 1.09\\
$M_{2i}$ ($M_{\odot}$) & 1.40 & 1.27 & 1.06 & 1.19 & 1.52\\
$M_{\text{lost}}$ ($M_{\odot}$) & 0.83 & 0.55 & 0.34 & 0.47 & 0.92\\
\hline
\end{tabular}
\end{center}
\label{conclusion-tab}
\end{table*}

\begin{table}
\renewcommand\arraystretch{1.2}
\centering
\caption{Summary statistics of relative temperature difference in contact binary systems.}
\begin{tabular}{lc}
\hline
\textbf{Statistic} & \textbf{Value} \\
\hline
Median DT\% & 3.29\% \\
90th percentile DT\% & 9.38\% \\
95th percentile DT\% & 13.41\% \\
Maximum DT\% & 48.83\% \\
Systems with DT\% $<$ 1\% & 17.3\% \\
Systems with DT\% $<$ 2\% & 32.9\% \\
Systems with DT\% $<$ 5\% & 66.1\% \\
Systems with DT\% $<$ 10\% & 91.1\% \\
Systems with DT\% $<$ 15\% & 95.3\% \\
Systems with DT\% $<$ 20\% & 96.6\% \\
\hline
\end{tabular}
\label{dt_stats}
\end{table}

\begin{figure*}
\centering
\includegraphics[width=0.44\textwidth]{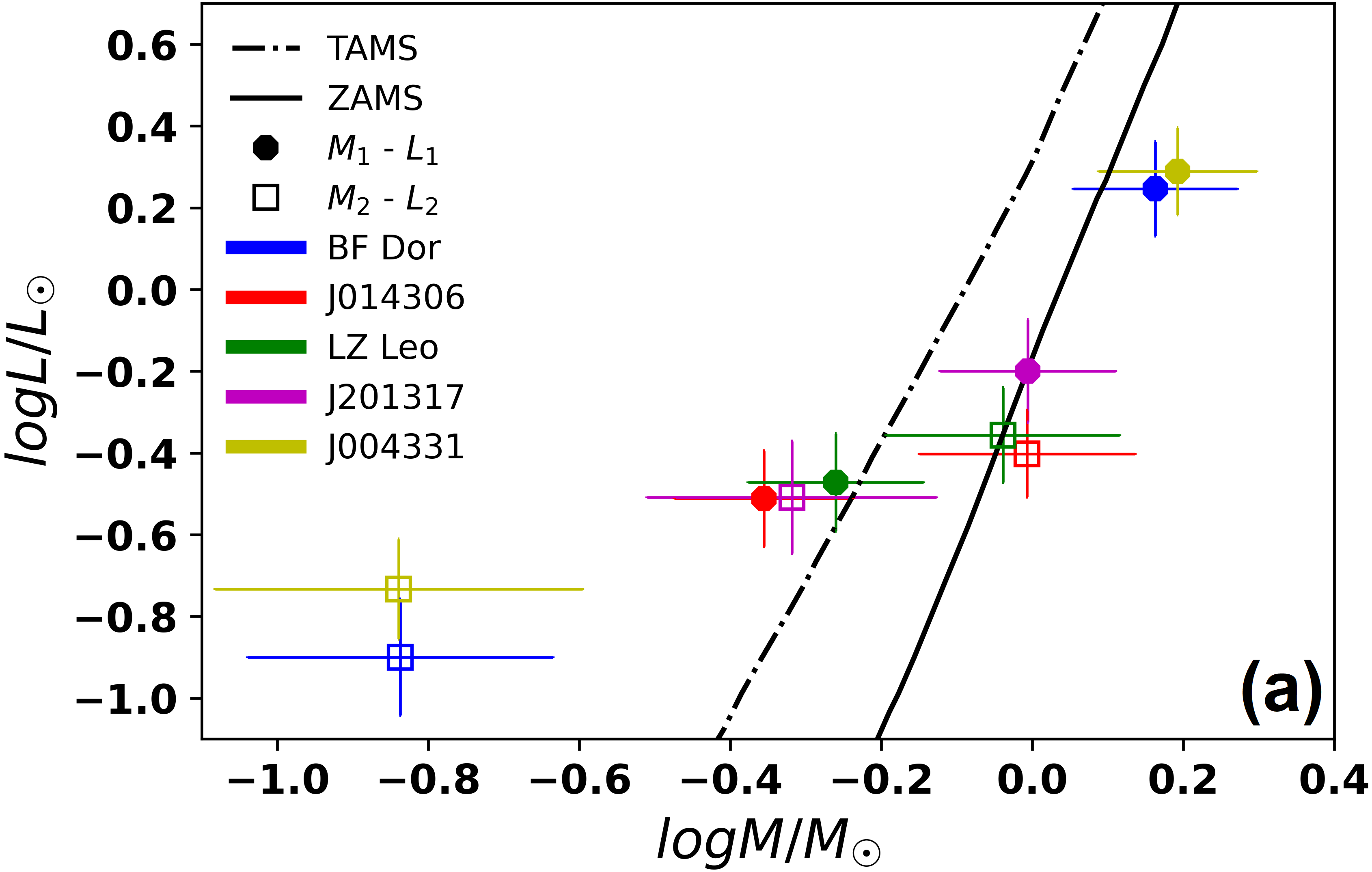}
\includegraphics[width=0.44\textwidth]{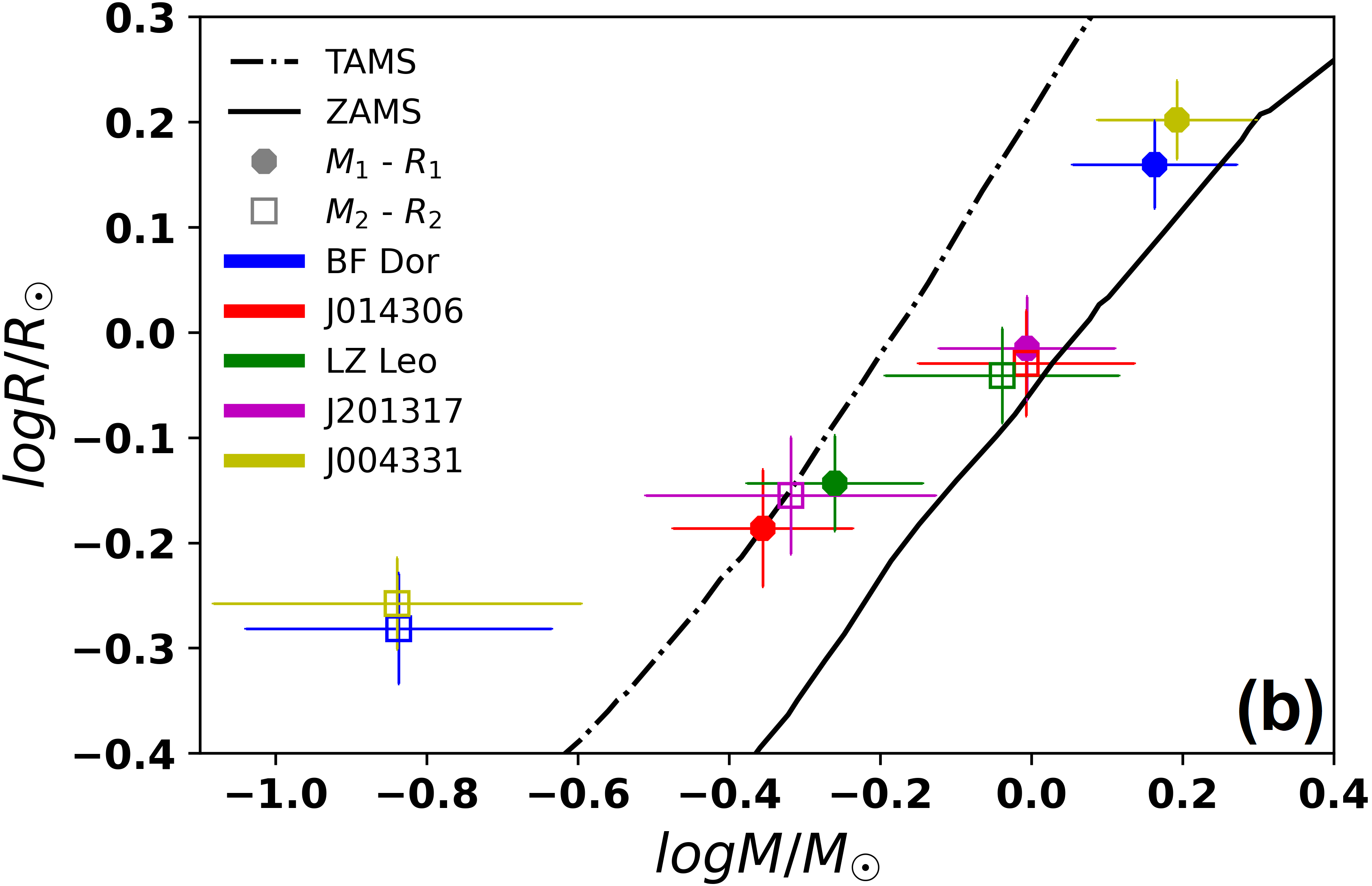}
\includegraphics[width=0.44\textwidth]{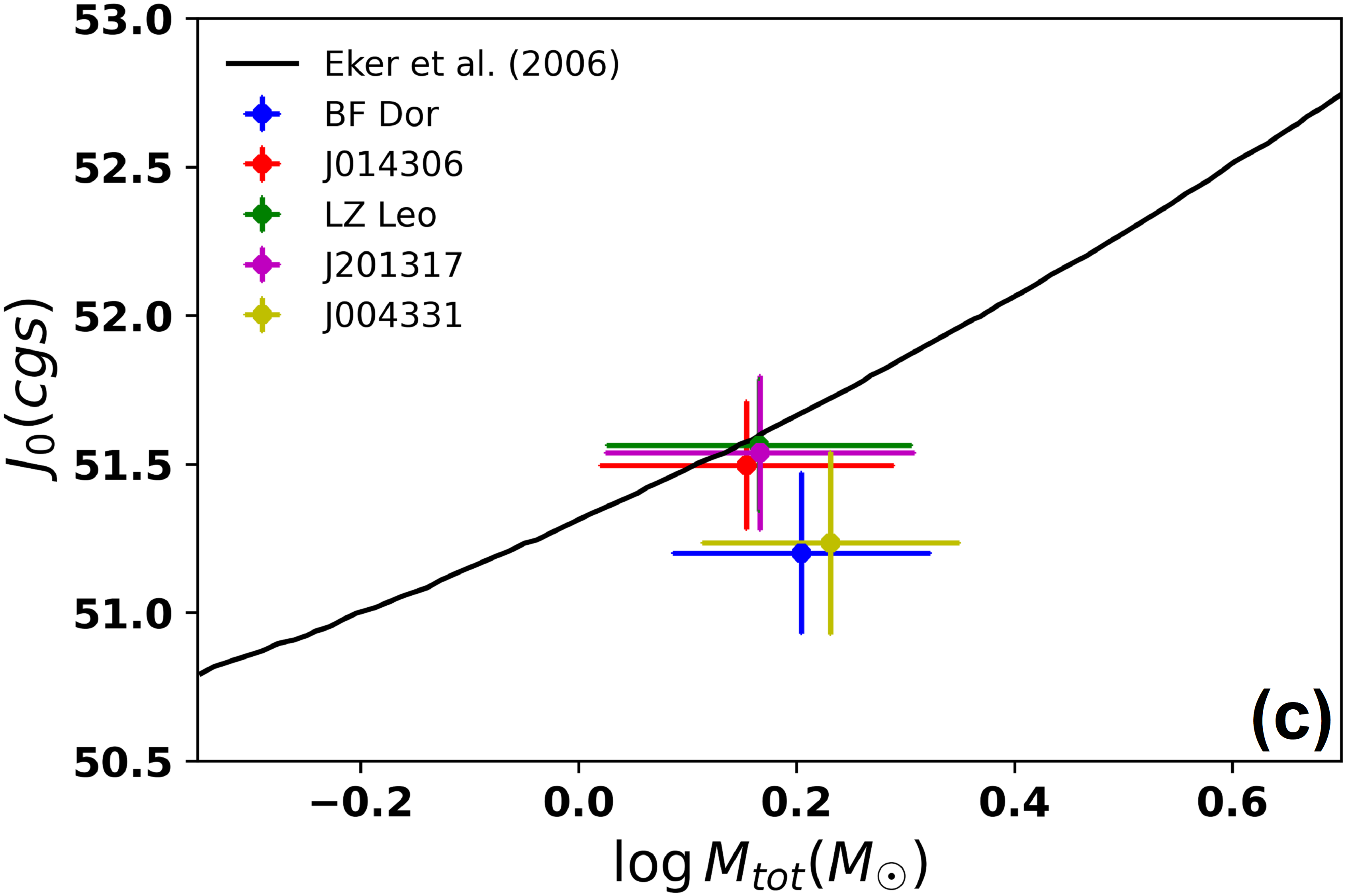}
\includegraphics[width=0.44\textwidth]{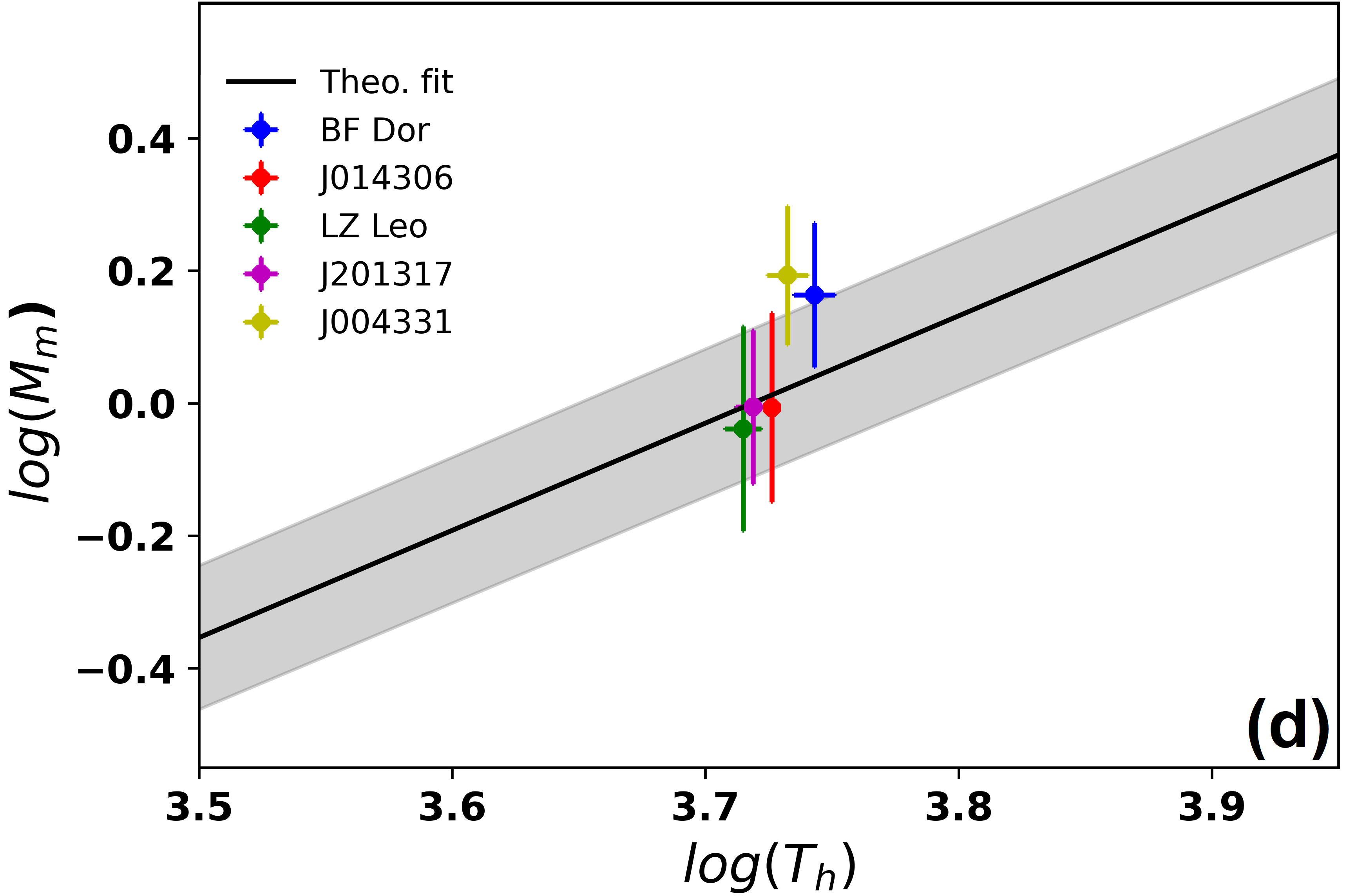}
\includegraphics[width=0.44\textwidth]{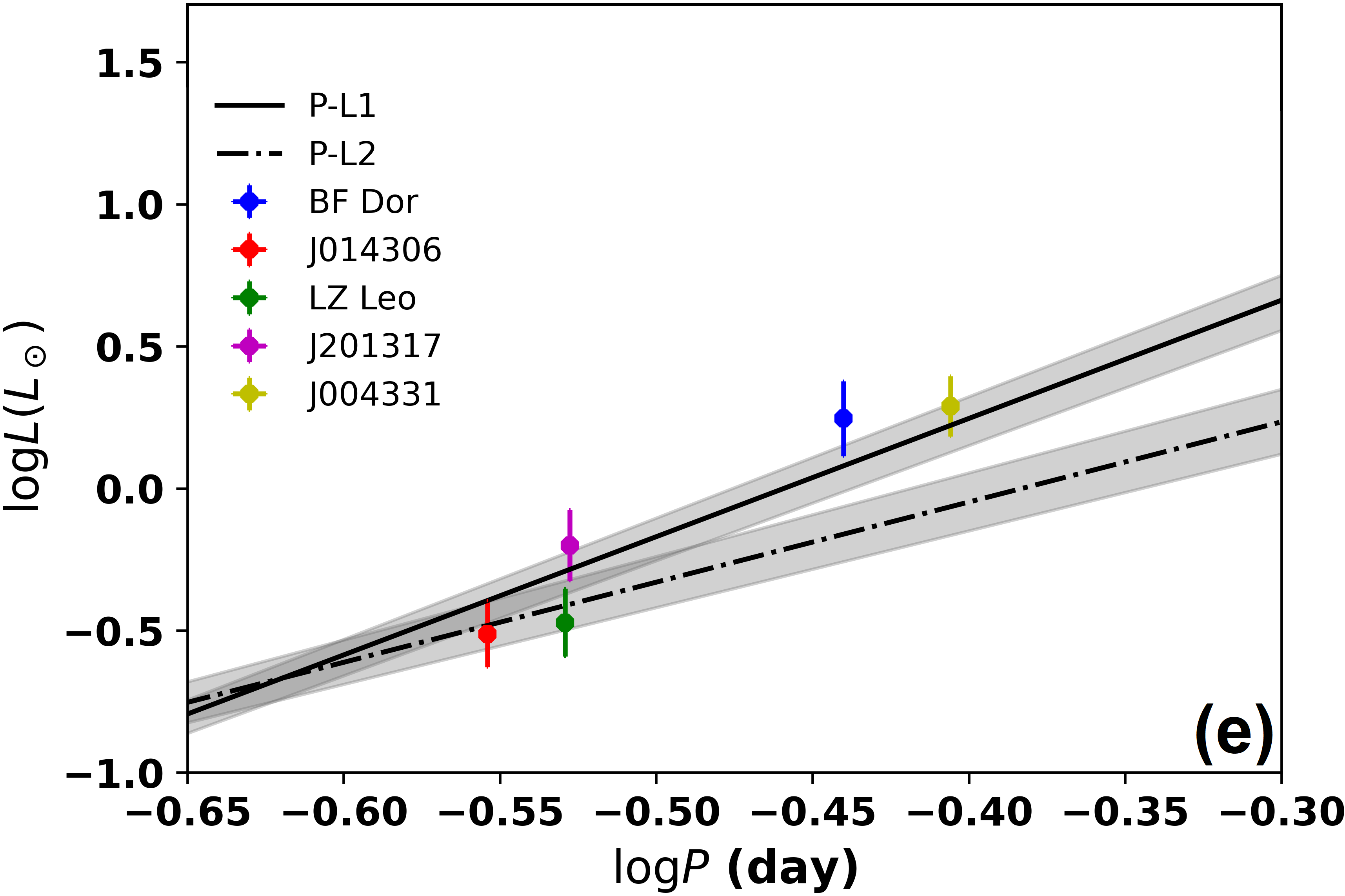}
\includegraphics[width=0.44\textwidth]{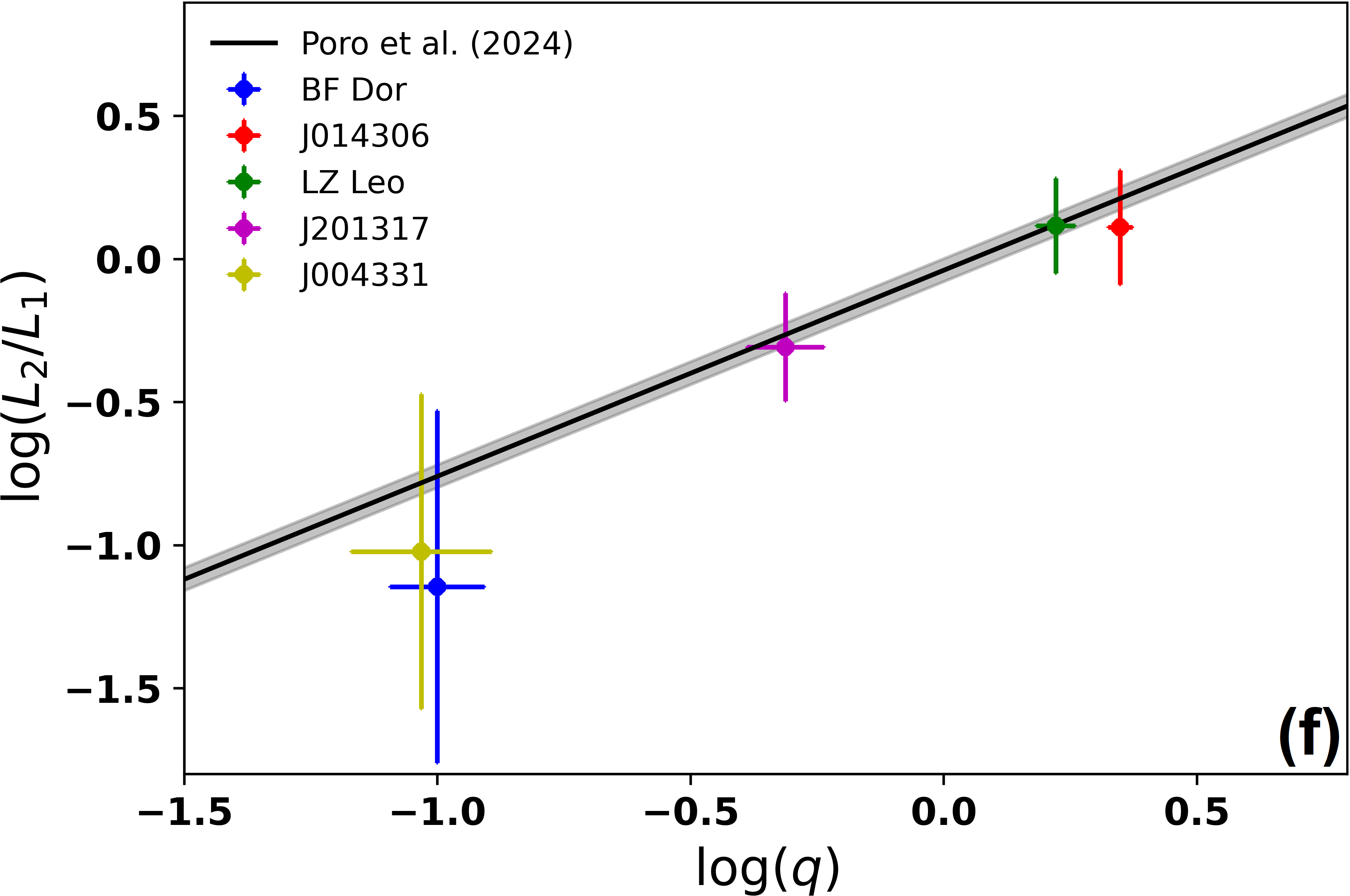}
\caption{a) $M-L$, b) $M-R$, c) $M_{\mathrm{tot}}$–$J_0$, d) $T_h-M_m$, e) $P-L$, f) $q-L_{ratio}$.}
\label{MRLetal}
\end{figure*}

C) To assess the evolutionary status of the targets, logarithmic Mass–Radius ($M$–$R$) and Mass–Luminosity ($M$–$L$) diagrams were presented based on the estimated absolute parameters (Table \ref{absolute}, Figure \ref{MRLetal}). The stellar components are plotted relative to the Zero-Age Main Sequence (ZAMS) and Terminal-Age Main Sequence (TAMS) as defined by \cite{2000AAS..141..371G}, adopting a metallicity of $Z=0.019$ (solar composition) along with the standard helium abundance and mixing-length parameters used in their models. As illustrated in Figure \ref{MRLetal}a,b, the more massive components in all systems lie above the TAMS line, while the less massive companions are positioned around the ZAMS line. It is important to note that contact binaries result from complex binary evolution and interaction mechanisms (\citealt{2005ApJ...629.1055Y}, \citealt{2011AcA....61..139S}), and their evolutionary paths diverge significantly from those of single stars. Consequently, any direct comparison with single-star ZAMS and TAMS lines should be approached with caution.

We used the $P$–$a$ empirical parameter relationship to estimate the absolute parameters of the target systems, and the system's total mass ($M_{\mathrm{tot}}$) was computed. Then, the orbital angular momentum of each system was estimated using Equation \ref{eqJ0}. We show the location of each system in the $\log M_{\mathrm{tot}}$–$J_0$ diagram (Figure \ref{MRLetal}c), based on the results in Table \ref{absolute}. The area below the quadratic line in Figure \ref{MRLetal}c is generally associated with contact binary stars, whereas the area above corresponds to detached systems \citep{2006MNRAS.373.1483E}, although this boundary is not strictly defined.

The study by \cite{2024RAA....24e5001P} used a sample of 428 contact binary systems to investigate an empirical $T_h$–$M_m$ relationship, where $T_h$ denotes the effective temperature of the hotter component and $M_m$ refers to the mass of the more massive star. Based on the mass derived from the estimated absolute parameters and the effective temperature obtained from the light curve solution, we show the position of the star on the $T_h$–$M_m$ diagram (Figure \ref{MRLetal}d). The component shows good agreement with the empirical relationship and its associated uncertainty.

The positions of the stars were examined with respect to two empirical parameter relationships proposed by \cite{2024RAA....24a5002P}: $P$–$L$ and $q$–$L_{\mathrm{ratio}}$. As shown in Figure \ref{MRLetal}e,f, the stars' locations show good agreement with the empirical fits and their uncertainties.

D) Based on the light curve solutions, which include the derived mass ratios, fillout factors, and orbital inclinations, the target systems are identified as eclipsing contact binaries. The subtype classification is determined by the effective temperature and mass of the components: if the less massive star has a higher temperature, the system is classified as W subtype; if the more massive star is hotter, it is classified as A subtype (\citealt{1970VA.....12..217B}). According to this criterion, three of the targets are classified as A subtype, while the remaining two belong to the W subtype (Table \ref{conclusion-tab}).

E) Determining the initial masses of the components in a contact binary system is essential for tracing their evolutionary paths. In this study, we applied the method proposed by \citet{2013MNRAS.430.2029Y}, which neglects the effects of energy transfer between the components.

To estimate the initial mass of the secondary component, we used an empirical relationship that incorporates its current mass and luminosity. The mass inferred from the luminosity, denoted as $M_L$, was calculated using a mass-luminosity relation (Equation \ref{eqML}). The mass gained by the secondary through accretion ($\Delta M$) was then added to its current mass ($M_2$) to derive the initial mass ($M_{2i}$) using Equation \ref{eqM1i}.

Then, the initial mass of the primary component ($M_{1i}$) was determined by subtracting the effective transferred mass from its current mass from Table \ref{absolute}, using Equation \ref{eqM2i}. This takes into account the mass lost from the system, modeled by the parameter $\gamma$, which expresses the ratio of lost mass to the total transferred mass. We adopted a value of $\gamma = 0.664$, as suggested by \cite{2013MNRAS.430.2029Y}. Additionally, a reciprocal mass ratio within the range $0 < 1/q < 1$ was used for consistency in the modeling.

The set of equations used in these calculations is given below:

\begin{equation}\label{eqM1i}
M_{2i} = M_2 + \Delta M = M_2 + 2.50(M_L - M_2 - 0.07)^{0.64},
\end{equation}

\begin{equation}\label{eqML}
M_L = \left(\frac{L_2}{1.49}\right)^{1/4.216},
\end{equation}

\begin{equation}\label{eqM2i}
M_{1i} = M_1 - (\Delta M - M_{lost}) = M_1 - \Delta M(1 - \gamma).
\end{equation}

Note that in above equations, $M_1$ and $M_2$ are the current masses of the primary and secondary components, respectively; $M_{1i}$ and $M_{2i}$ represent their initial masses. The term $M_L$ corresponds to the mass estimated from the secondary's luminosity, while $\Delta M$ is the transferred mass, and $M_{lost}$ refers to the mass expelled from the system. The resulting initial masses, listed in Table \ref{conclusion-tab}, show good agreement with the studies by \citet{2013MNRAS.430.2029Y} and \citet{2014MNRAS.437..185Y}.

F) Contact binaries with extremely low mass ratios present significant observational and theoretical challenges. The detection of reliable radial velocity curves is often difficult or even impossible due to the extremely faint and low-mass secondaries, compounded by the broad and blended spectral lines (\citealt{2019AJ....158..186K}). In such cases, high-quality photometric light curves become especially valuable for deriving physical parameters and analyzing the nature of these systems. These binaries are of particular interest, as they are considered likely progenitors of stellar mergers and offer key insights into the physics of binary coalescence.

Numerous theoretical studies have proposed that contact binaries possess a lower-limit cut-off in mass ratio, below which they become dynamically unstable and may ultimately merge. An early estimate by \cite{1995ApJ...444L..41R}, which ignored the spin angular momentum of the secondary, placed this limit at approximately \( q_{\mathrm{min}} \sim 0.09 \). Subsequent models that included additional angular momentum effects have suggested lower thresholds: \cite{2006MNRAS.369.2001L} proposed \( q_{\mathrm{min}} \sim 0.076\text{--}0.078 \); \cite{2007MNRAS.377.1635A} and \cite{2009MNRAS.394..501A} found \( q_{\mathrm{min}} \sim 0.070\text{--}0.109 \); and \cite{2010MNRAS.405.2485J} estimated \( q_{\mathrm{min}} \sim 0.05\text{--}0.105 \), depending on the internal structure of the primary. A statistical analysis by \cite{2015AJ....150...69Y} yielded \( q_{\mathrm{min}} \approx 0.044 \), based on the relations between \( q{-}f \) and \( q{-}J_{\mathrm{spin}}/J_{\mathrm{orb}} \). More recently, \cite{2024NatSR..1413011Z} and \cite{2024A&A...692L...4L} proposed even lower thresholds of \( q_{\mathrm{min}} \approx 0.038\text{--}0.041 \) and \( q_{\mathrm{min}} = 0.0356 \), respectively.

BF Dor and J004331 target systems are classified as contact binaries with extremely low mass ratios, estimated to be 0.10 for BF Dor and 0.093 for J004331 in this study. These values place them very close to the theoretical threshold for orbital instability, below which contact binaries are expected to become dynamically unstable and eventually merge. The proximity of BF Dor and J004331 to this limit suggests that they may be in advanced evolutionary stages, making them important targets for testing predictions of binary instability and merger scenarios.

Assessing the dynamical stability of the contact binary systems BF Dor and J004331 requires evaluating the ratio of spin angular momentum ($J_{\mathrm{spin}}$) to orbital angular momentum, as described by \citet{hut1980stability}. In this study, we employed the equation provided by \citet{2015AJ....150...69Y} to compute $J_{\mathrm{spin}}/J_0$ for the two systems analyzed:

\begin{equation}\label{eqJspin}
\frac{J_{\mathrm{spin}}}{J_0} = \frac{1 + q}{q} \left[ (k_1 r_1)^2 + (k_2 r_2)^2 q \right],
\end{equation}

\noindent where, $k_1$ and $k_2$ denote the dimensionless gyration radii, and $r_1$ and $r_2$ represent the relative radii of the components. The adopted values for $k_1$ and $k_2$ are taken from \citet{2006MNRAS.369.2001L}. We found that the values of $J_{\mathrm{spin}}/J_0$ are 0.0193 for BF Dor and 0.0207 for J004331, suggesting that both systems are dynamically stable in terms of their $J_{\mathrm{spin}}/J_0$ ratios \citep{2006MNRAS.369.2001L}. According to the study by \citet{wadhwa2021zz}, however, our calculations imply that both systems would be dynamically unstable if their instability mass ratios were smaller than $\approx 0.075$ for BF Dor and $\approx 0.070$ for J004331.

\vspace{0.6cm}
\section*{Data availability}
Ground-based data are available in the paper's online supplement.

\vspace{0.6cm}
\section*{Acknowledgments}
This manuscript, including the observation, analysis, and writing processes, was provided by the BSN project (\url{https://bsnp.info/}). Two binary systems' data in this work is based on observations carried out at the Observatorio Astron\'omico Nacional on the Sierra San Pedro M\'artir which is operated by the Universidad Nacional Aut\'onoma de M\'exico. We used IRAF, distributed by the National Optical Observatories and operated by the Association of Universities for Research in Astronomy, Inc., under a cooperative agreement with the National Science Foundation. We used data from the European Space Agency mission Gaia (\url{http://www.cosmos.esa.int/gaia}). This study includes data from the TESS mission, which is funded by NASA's Explorer Program. The authors would like to thank Ehsan Paki for his valuable support and insightful discussions that contributed to the development of this work.

\vspace{0.6cm}
\bibliography{REFS}{}
\bibliographystyle{aasjournal}

\end{document}